\documentclass{emulateapj}
\usepackage{graphicx}
\usepackage{apjfonts}
 
\usepackage{bm}
\slugcomment{\scriptsize{Accepted for Publication in ApJ, Preprint typeset using \LaTeX style}} 
\shorttitle{Effects of Penetrative Convection on Solar Dynamo}
\shortauthors{Masada, Yamada and Kageyama}
\begin{document}
\title{Effects of Penetrative Convection on Solar Dynamo} 
\author{Youhei Masada\altaffilmark{1}, Kohei Yamada\altaffilmark{1} and Akira Kageyama \altaffilmark{1}} 
\altaffiltext{1}{Department of Computational Science, Graduate School of System Informatics, Kobe University; 
Kobe, Japan, contact: ymasada@harbor.kobe-u.ac.jp} 

\begin{abstract}
Spherical solar dynamo simulations are performed. Self-consistent, fully compressible magnetohydrodynamic system with a stably stratified 
layer below the  convective envelope is numerically solved with a newly developed simulation code based on the Yin-Yang grid. The effects 
of penetrative convection are studied by comparing two models with and without the stable layer. The differential rotation profile in both models 
is reasonably solar-like with equatorial acceleration. When considering the penetrative convection, a tachocline-like shear layer is 
developed and maintained beneath the convection zone without assuming any forcing. While turbulent magnetic field becomes predominant 
in the region where the convective motion is vigorous, mean-field component is preferentially organized in the region where the convective 
motion is less vigorous. Especially in the stable layer, the strong large-scale field with a dipole symmetry is spontaneously built up. 
The polarity reversal of the mean-field component takes place globally and synchronously throughout the system regardless the presence of 
the stable layer. Our results suggest that the stably stratified layer is a key component for organizing the large-scale strong magnetic field, but 
is not essential for the polarity reversal. 
\end{abstract}
\keywords{convection--magnetohydrodynamics (MHD) -- Sun: dynamo -- Sun: interior}
\section{Introduction}
A grand challenge in the solar physics is a construction of self-consistent theory that explains the observed large-scale spatial structures 
of the fields and their dynamical change in time. Two basic large-scale structures that are left to be explained are the azimuthal average 
of the azimuthal flow, $\bar{v}_\phi$, and the azimuthal average of the azimuthal magnetic field, $\bar{B}_\phi$. The averaged velocity 
$\bar{v}_\phi$ is characterized by the conical iso-rotation profile in the meridian plane and the thin tachocline layer with steep angular velocity 
gradient (e.g., \citealt{thompson+03}). The averaged magnetic field $\bar{B}_\phi$ is characterized by antisymmetric profile with respect 
to the equator and the polarity reversals with the pseudo-periodicity of $22$ years 
(e.g., \citealt{hathaway10}). See \citet{ossendrijver03} and \citet{miesch05,miesch12} for reviews.

To reproduce the large-scale structures and dynamics, magnetohydrodynamics (MHD) simulations have been performed both in the global 
(spherical shell) geometry (e.g., \citealt{gilman+81,gilman83,glatzmaier85}) and in the local Cartesian geometry 
(e.g., \citealt{meneguzzi+89,cattaneo+91,nordlund+92, brandenburg+96}). 

The first modern solar dynamo simulation with solar values of luminosity, background stratification, and rotation rate was performed by 
\citet{brun+04}. They solved anelastic MHD convection system in the domain that extends over 
$0.72$--$0.97R_\odot$, spanning the bulk of the convection zone. While the solar-like equatorial acceleration and the dynamo-generated magnetic 
field with strengths of order $5000$ G was achieved, the mean large-scale magnetic field were relatively weak and did not exhibit periodic polarity 
reversals. 

\citet{browning+06} showed, in anelastic spherical shell dynamo simulation with the presence of the tachocline, that strong axisymmetric toroidal 
magnetic fields can be formed in stably stratified layer below the convection zone. The associated mean poloidal magnetic fields 
showed the dipole dominance, but they did not exhibit polarity reversals. While the solar-like rotation profile was 
achieved in their simulations, a mechanical forcing was necessary to maintain the thin tachocline layer with steep angular velocity gradient.  

Solar dynamo simulations that successfully produced the cyclic large-scale magnetic fields were presented in \citet{ghizaru+10} and 
\citet{racine+11}. Their simulations are based on an anelastic model that is commonly used in the global circulation models of the earth's 
atmosphere with a cooling term to force the system toward the ambient state (e.g., \citealt{prusa+08,smolarkiewicz09}). The solar-like thin 
tachocline layer was developed as a consequence of the cooling as well as the low dissipation embodied in their numerical scheme. They showed 
that the large-scale magnetic field is built up in the tachocline layer and exhibits polarity reversals when the temporal integration of the 
simulation was calculated for long enough. 

The large-scale dynamo activity was found not only in the anelasic models but also in the compressible dynamo simulation. \citet{kapyla+10} 
performed the dynamo simulation with the penetrative convection in a spherical-wedge geometry (e.g., \citealt{brandenburg+07}). 
Using a weakly stratified dynamo model, they succeeded to simulate the formation and the cyclic polarity reversal of the 
large-scale magnetic field. Unlike \citet{browning+06} and \citet{ghizaru+10}, the large-scale dynamo operated in the convection zone in 
their model. Despite the presence of the underlying stable layer below the convective envelope, the spontaneous formation of the solar-like 
tachocline layer was not observed. 

These numerical studies that targeted for the solar dynamo have made it increasingly clear that the underlying stable layer below the convection zone 
is an important building block for the solar dynamo. It seems to play a crucial role in the formation of the solar-like $\bar{v}_\phi$ \& $\bar{B}_\phi$. 
However, there is no research that directly compares two dynamo simulations differing only in the presence and absence of the underlying 
stable layer. 

In \citet{miesch+09}, the influence of the tachocline on the magnetic dynamo was reviewed by comparing two previous simulations done 
by \citet{brun+04} and \citet{browning+06}. While two simulation models are both based on the same simulation code with solar values 
of the luminosity, rotation rate, and background stratification, they adopt different diffusivities and grid spacings that can affect the convective 
motion and magnetic dynamo. To get a better grasp of the role of the stably stratified layer in the solar dynamo mechanism, the influences of 
other parameters than the presence of the stable layer should be eliminated. 
This is one of motivations of our study. 

In this paper, we perform fully compressible spherical solar dynamo simulation with a stably stratified layer below the convection zone. 
Formations of the key profiles of the solar interior,  i.e., the solar-like $\bar{v}_\phi$ \& $\bar{B}_\phi$, and the spontaneous polarity 
reversals are reproduced without assuming any forcing in the fundamental equations. To elucidate the effects of the penetrative convection, 
two simulations with and without the stable layer below the convection zone are compared. 

Another purpose of this paper is to report a development of new program code for the solar dynamo simulation. A lot of simulation models for the 
global dynamo are spectral-based type, using the spherical harmonics expansion (e.g., \citealt{brun+04}). The spherical harmonics 
expansion method is, however, believed to be confronted with the parallel scaling difficulty when tens of thousands, or more, processor cores 
are used. A different approach to massively parallel solar dynamo model is strongly required for the present peta- or coming exa-scale era. We 
have developed a global solar dynamo simulation code based on the grid point-based approach.

The spherical geometry imposes difficulties in the design of the spatial grid points to sustain high numerical efficiency, accuracy, and 
parallel scalability. We have proposed an overset grid method approach to the spherical geometry~(\citealt{kageyama+04}). The grid 
system, Yin-Yang grid, is applied to geodynamo (e.g., \citealt{kageyama+08,miyagoshi+10}), mantle convection (e.g., \citealt{kameyama+08,
tackley08}), supernova explosions (e.g., \citealt{muller+12,lentz+12}), and other astro- and geophysical simulations. The parallel scaling property 
of the spherical MHD simulation on the Yin-Yang grid is promising. It attained $46$\%  ($15.2$ TFlops) of the peak performance 
of $4096$ cores of the Earth Simulator supercomputer for the geodynamo simulations (Gordon Bell Award in Supercomputing 2004). 
Our new solar dynamo code is developed based on this Yin-Yang geodynamo code. This paper is our first report on the results obtained 
by this Yin-Yang solar dynamo code. 

\section{Numerical Settings}
\begin{figure}[tpb]
\begin{center}
\scalebox{0.53}{{\includegraphics{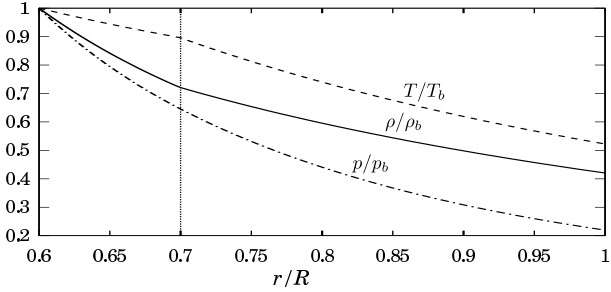}}} 
\caption{Radial profiles of the initial temperature (solid), initial density (dashed), and initial pressure (dash-dotted) adopted in our dynamo 
model. The vertical axis is normalized by their values at $r=0.6R$.}
\label{fig1}
\end{center}
\end{figure}
\begin{figure}[tpb]
\begin{center}
\scalebox{0.4}{{\includegraphics{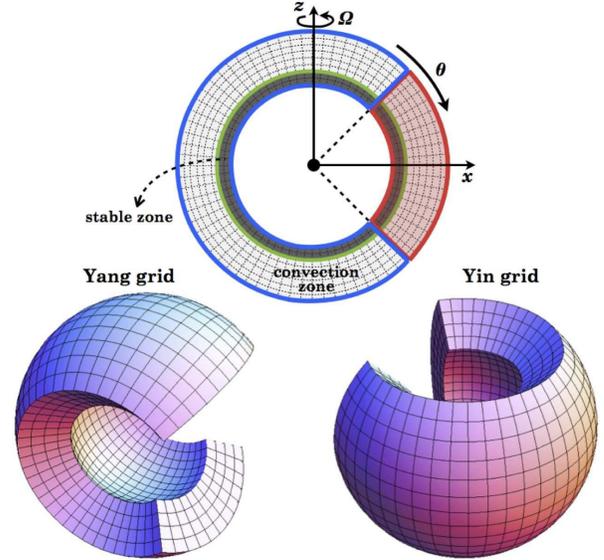}}} 
\caption{Yin-Yang grid. Each of the two congruent grids, Yin-grid and Yang-grid, covers a partial spherical shell region defined as 
($\pi/4 \le \theta \le 3\pi/4 , -3\pi/4 \le \phi \le 3\pi/4$). They are combined in a complemental way to cover a whole spherical shell. 
The domains surrounded by red and blue curves are assigned to the Yin and Yang grids, respectively.  }
\label{fig2}
\end{center}
\end{figure}
We numerically solve an MHD dynamo convection system in a spherical shell domain defined by $(0.6R\le r \le R)$, $(0\le \theta \le \pi)$, 
and $(-\pi \le \phi < \pi)$, where $r$, $\theta,$ and $\phi$ are radius, colatitude, and longitude, respectively. 
Our model has two layers: upper convective layer of thickness $0.3R$ in the range of $(0.7R \le r \le R)$, and stably stratified lower layer 
of thickness $0.1R$ in $(0.6R \le r \le 0.7R)$. 

The fundamental equations are the fully compressible MHD equations in the rotating frame of reference with a constant angular velocity 
$\bm{\Omega} = \Omega_0 \bm{e}_z$ which is parallel to the coordinate axis ($\theta=0$):
\begin{eqnarray}
\frac{\partial \rho}{\partial t} & = & - \nabla\cdot \bm{f}   \;, \label{eq1} \\ 
\frac{{\partial }\bm{f}}{ \partial  t} & = & - \nabla \cdot (\bm{v} \bm{f})  - \nabla p  + \bm{j}\times\bm{B}  \nonumber \\
&& + \rho \bm{g} + 2\rho \bm{v} \times \bm{\Omega} +  \mu \left[ \nabla^2 \bm{v} + \frac{1}{3}\nabla (\nabla\cdot \bm{v})\right]  \;, \ \ \ \ \label{eq2} \\
\frac{\partial p }{\partial t} & = &  - \bm{v}\cdot\nabla p  - \gamma p \nabla\cdot \bm{v} \nonumber \\
&&+ (\gamma-1)\left[ \nabla \cdot (\kappa \nabla T) + \eta \bm{j}^2 +\Phi \right] \;, \label{eq3} \\
\frac{\partial \bm{A} }{\partial t} & = & \bm{v}\times\bm{B} - \eta \bm{j}\;, \label{eq4}
\end{eqnarray}
with 
\begin{eqnarray}
&& \Phi = 2\mu\left[ e_{ij}e_{ij} - \frac{1}{3} \left( \nabla \cdot  \bm{v} \right) \right] \;,
e_{ij} = \frac{1}{2}\left( \frac{\partial v_i}{\partial x_j} + \frac{\partial v_j}{\partial x_i}\right) \;, \nonumber \\
&& \bm{g} = -g_0/r^2\bm{e}_r\;,\  \bm{B} = \nabla \times \bm{A} \;,\ \bm{j} = \nabla\times \bm{B} \;.\nonumber
\end{eqnarray}
Here the mass density $\rho$, pressure $p$, mass flux $\bm{f} = \rho\bm{v}$, magnetic field's vector potential $\bm{A}$ are the basic 
variables. We assume an ideal gas law $p = (\gamma -1)\epsilon$ with $\gamma = 5/3$, where $\epsilon$ is the internal energy. 
The viscosity, electrical resistivity, and thermal conductivity are represented by $\mu$,  $\eta$, and $\kappa$ respectively.

The initial condition is a hydrostatic equilibrium which is described by a piecewise polytropic distribution with the polytropic index $m$,
\begin{equation}
\frac{dT}{dr} = \frac{g_0}{c_v(\gamma-1)(m + 1)}\;, 
\end{equation}
(e.g., \citealt{kapyla+10}). We choose $m=1$ and $3$ for the upper convection layer and the lower stable layer, respectively. 
The thermal conductivity is determined by requiring a constant luminosity $L$, defined by $L \equiv -4\pi\kappa r^2d T/d r$,   
throughout the domain. 

We solve the MHD equations in a non-dimensional form.  
Normalization quantities are defined by setting $R = g_0=\rho_0=1$ where $\rho_0$ is the initial density at  $r=0.6R$. 
We normalize length, time, velocity, density, and magnetic field in units of $R$, $\sqrt{R^3/g_0}$, $\sqrt{g_0/R}$, 
$\rho_0$ and $\sqrt{g_0\rho_0/R}$. We define the Prandtl, magnetic Prandtl, and Rayleigh numbers by
\begin{equation}
{\rm Pr}  =  \frac{\mu}{\kappa},\ \ {\rm Pm} = \frac{\mu}{\eta},\ \ 
 {\rm Ra}  =  \frac{GMd^4\rho_m^2}{\mu\kappa R^2} \left(- \frac{{\rm d} s }{{\rm d}r}\right)_{r_m} \;, 
\end{equation}
where $\rho_{m}$ is the density at the mid-convection zone ($r=r_m$), and $d = 0.3R$ 
is the depth of the convection zone. The stratification level is controlled by the normalized pressure scale height at the surface, 
\begin{equation}
\xi_{0} \equiv \frac{c_v(\gamma-1)T_s}{g_0 R}\;,
\end{equation}
where $T_s$ is the temperature at $r=R$. In this work, we use $\xi_0 = 0.3$, yielding 
a small density contrast about $3$. Figure~1 shows the radial distributions of the initial temperature, density and pressure adopted for 
our numerical model by solid, dashed and dash-dotted curves, respectively. The vertical axis is normalized by the value at $r=0.6R$. 
The radial slopes in our numerical model are more gentle than the solar profiles. These give the convective motion with the Mach number of 
$\mathcal{O}(10^{-2})$. 
 
The relative importance of rotation in the convection is measured by the Coriolis number 
\begin{equation}
{\rm Co} = \frac{2\Omega_0d}{v_{\rm rms}} \;,
\end{equation}
where $v_{\rm rms} \equiv \langle \langle v_\theta^2 + v_r^2 \rangle \rangle^{1/2}$ 
is the mean velocity. The double angular brackets  denote the time and volume average in the convection zone 
in the saturated state. The convective turn-over time and the equipartition strength of magnetic field are defined, respectively, by 
\begin{equation}
\tau_c \equiv \frac{d}{v_{\rm rms }} \;, \ \ \ B_{\rm eq} \equiv \langle\langle \rho\; (v_\theta^2 + v_r^2) \rangle\rangle^{1/2} \;.
\end{equation} 

The stress-free boundary condition for the velocity is imposed on the two spherical boundaries. We assume the perfect conductor boundary 
condition for the magnetic field ($A_r = A_\theta = A_\phi = 0$) on the inner surface, and the radial field condition 
($A_r = 0, \partial A_\theta/\partial r = -A_\theta/r, \partial A_\phi/\partial r = -A_\phi/r$) on the outer surface.
A constant energy flux is imposed on the inner boundary. The temperature is fixed to be $T_s$ on the outer boundary. 

The eqs.~(\ref{eq1})--(\ref{eq4}) are discretized by the second-order central difference on the Yin-Yang grid. Each of the two congruent 
grids, Yin-grid and Yang-grid, covers a partial spherical shell region defined as ($\pi/4 \le \theta \le 3\pi/4 , -3\pi/4 \le \phi \le 3\pi/4$).
They are combined in a complemental way to cover a whole spherical shell as shown in Figure~2. The regions surrounded by red and 
blue curves are assigned to the Yin- and Yang-grids, respectively. Physical quantities on the horizontal boarders of Yin- or Yang-grid are 
set by mutual interpolations. For the time integration, the standard fourth-order Runge-Kutta method is used.  Since the Yin-Yang grid is 
free from the coordinate singularity and the grid concentration around there, we can avoid the severe time-step constraint due to the CFL 
condition. See \citealt{kageyama+04} for details on the Yin-Yang grid method. The computation is performed in parallel using MPI 
(Message Passing Interface). 

Non-dimensional parameters $\rm{Pr}=0.2$, $\rm{Pm}=4.0$, and $\rm{Ra} = 1.2\times10^5$, and constant angular velocity 
of $\Omega_0 = 0.4$ are adopted in all the calculations reported here in order to achieve the Coriolis number expected in the convection 
zone of the Sun [${\rm Co} \simeq \mathcal{O}(1)$]. The total grid size for the run with the upper convection layer and the lower stable layer 
(Model A) is $121$ (in $r$) $\times 402$ (in $\theta$) $\times$ 402 (in $\phi$) $\times 2$ (Yin \& Yang). A model without the stable layer 
(Model B) is also studied, in the domain $(0.7R \le r \le R)$, with the same physical parameters and the same grid spacings 
($91 \times 402 \times 402  \times 2$). A random temperature perturbation and weak magnetic field are seeded in the convection zone 
when the calculation starts. 
\section{Numerical Results}
\begin{figure}[tbp]
\begin{center}
\scalebox{0.45}{{\includegraphics{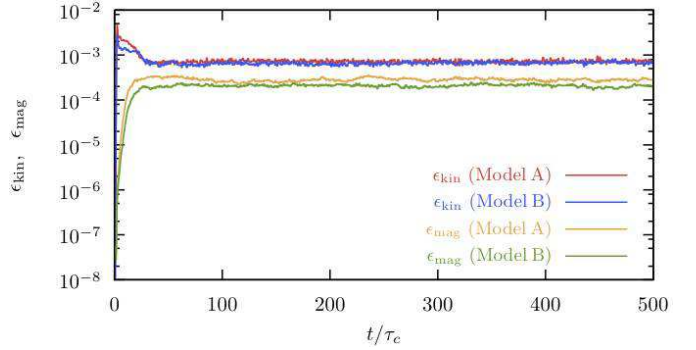}}} 
\caption{Temporal evolutions of volume-averaged kinetic and magnetic energies ($\epsilon_{\rm kin}$ and $\epsilon_{\rm mag}$) 
for Models~A and~B. The red and orange curves denote $\epsilon_{\rm kin}$ and $\epsilon_{\rm mag}$ for Model~A, and the blue 
and green curves are those for Model~B.}
\label{fig3}
\end{center}
\end{figure}
\begin{figure*}[tbp]
\begin{center}
\scalebox{0.9}{{\includegraphics{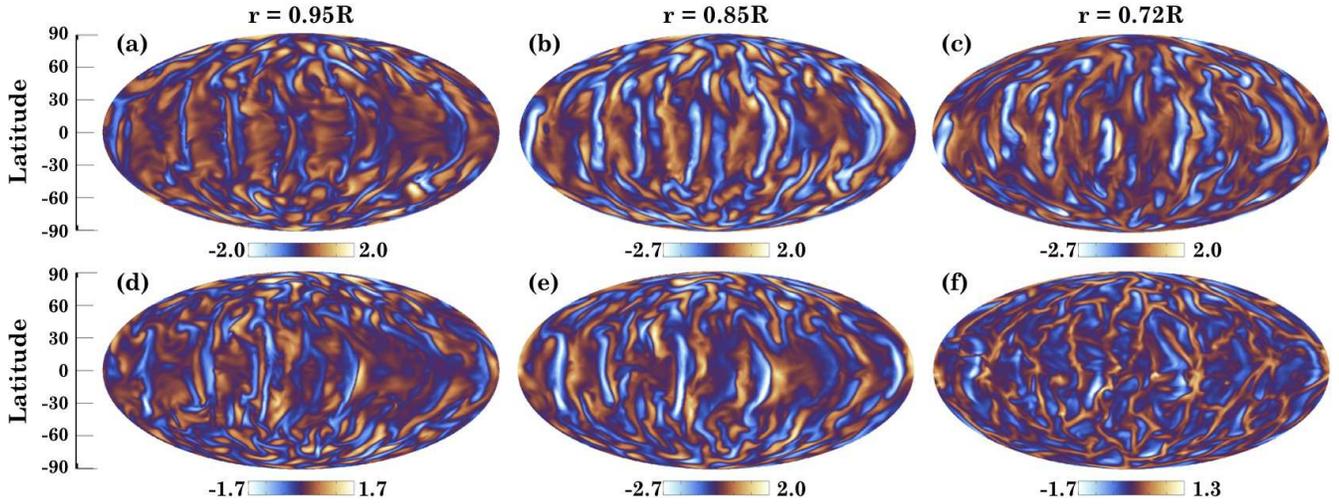}}} 
\caption{Distribution of radial velocity on spherical surfaces at sampled radii $v_r (\theta,\phi)$ when $t=330\tau_c$ (in the Mollweide 
projection). Panels~(a)--(c) correspond to the radii $r=0.95R$, $0.85R$ and $0.72R$ for Model~A, and panels~(d)--(f) 
are those for Model~B. The orange and blue tones depict upflow and downflow velocities normalized by $v_{\rm rms} = 0.03$.}
\label{fig4}
\end{center}
\end{figure*}
\begin{figure}[tbp]
\begin{center}
\scalebox{1.1}{{\includegraphics{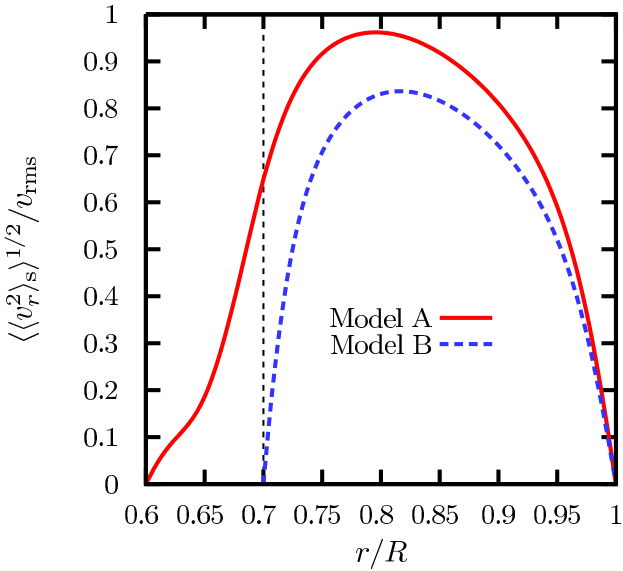}}} 
\caption{Radial profile of the mean radial velocity $\langle\langle v_r^2 \rangle_{\rm s}\rangle^{1/2}$. The time average spans 
in the range of $300\tau_c \le t \le 400 \tau_c$. The vertical axis is normalized by $v_{\rm rms} = 0.03$. The red-solid and 
blue-dashed curves correspond to the models~A and~B, respectively. The vertical dashed line denotes the base of the convection zone. }
\label{fig5}
\end{center}
\end{figure}
\begin{figure*}[tbp]
\begin{center}
\scalebox{0.9}{{\includegraphics{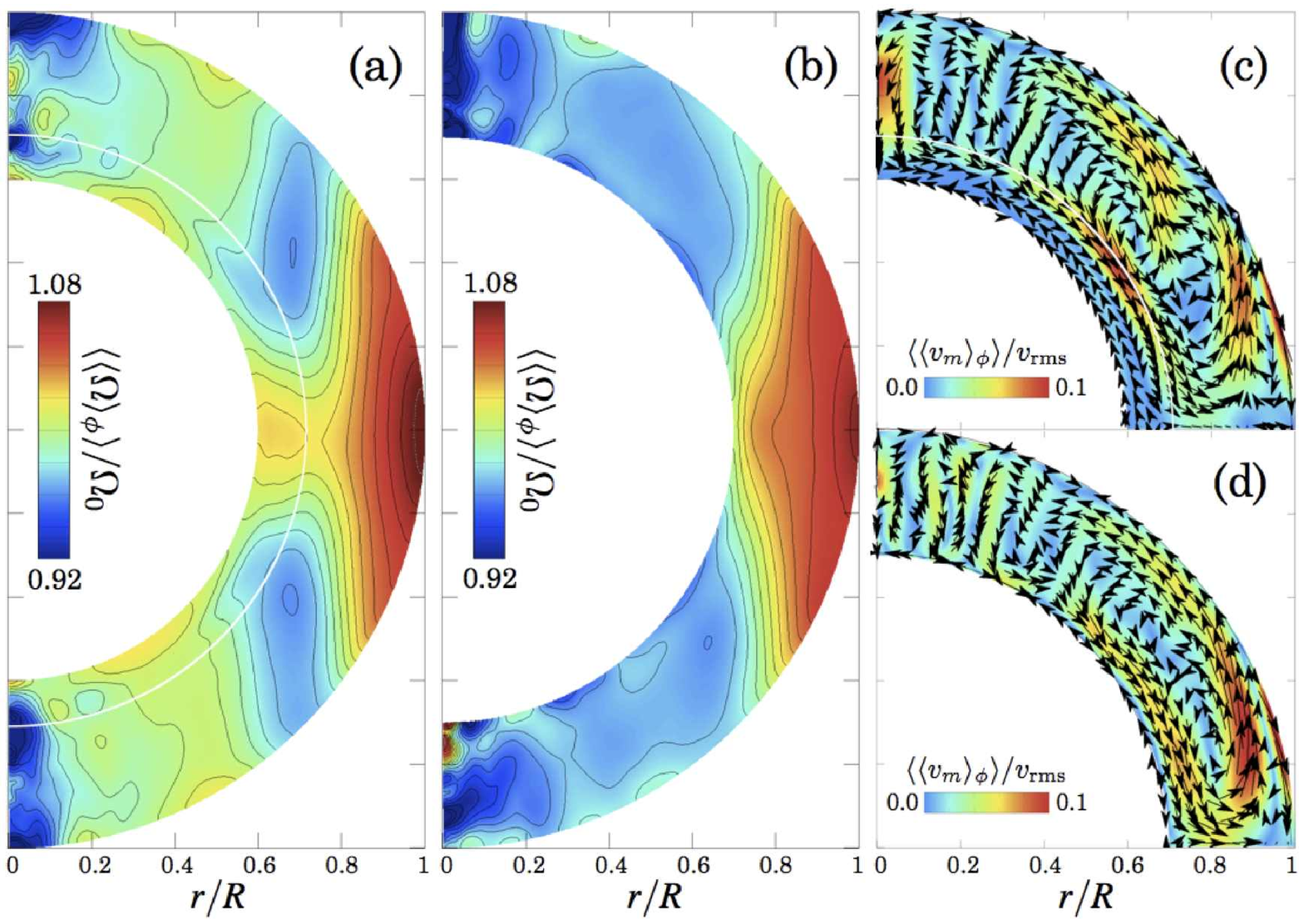}}} 
\caption{Mean angular velocity $\langle\langle \Omega \rangle_\phi \rangle $ [panels~(a) and (b) for Models~A and B], 
and mean meridional flow  [panels~(c) and (d) for Models~A and~B], where 
$\langle \langle \Omega \rangle_\phi \rangle  = \langle\langle v_{\phi} \rangle_\phi \rangle /(r \sin\theta) + \Omega_0 $. The 
mean meridional flow velocity is defined by  $\langle\langle v_m \rangle_\phi \rangle = [\langle\langle v_r \rangle_\phi \rangle^2 + \langle\langle v_\theta \rangle_\phi \rangle^2]^{1/2}$. 
The white solid curves in panels~(a) and~(c) denote the interface between the convective and stable layers.}
\label{fig6}
\end{center}
\end{figure*}
Figure~3 shows the temporal evolution of the volume-averaged kinetic and magnetic energies defined by 
\begin{equation}
\epsilon_{\rm kin} = \int \frac{1}{2}\rho {\bf v}^2 dV \Big / \int dV \;, \   \epsilon_{\rm mag} \equiv \int \frac{\bf B^2}{2\mu_0}\Big / \int dV \;,  
\end{equation}
for Models~A and~B. The red and orange curves correspond to those for Model~A. The blue and green curves are for Model~B. After the 
convective motion sets in, it reaches a nonlinear saturation state at around $t = 30\tau_c$. The saturation 
levels of the convection kinetic energy for Models~A and~B are almost the same. The mean velocity is $v_{\rm rms} = 0.03$ which yields 
$B_{\rm eq} = 0.02$, $\rm{Co} = 8.0$ and $\tau_c = 10.0$ for both models. We have run the simulations till $500\tau_c$ and then compare 
physical properties of convections, mean flows and magnetic dynamos between two models. 

To examine the convective and magnetic structures in detail, we define the following four averages of a function $h(\theta,\phi)$ on a sphere.\\
\\
The latitudinal average: 
\begin{equation}
\langle h\rangle_\theta \equiv \frac{1}{2}\int_{-1}^1 h(\theta,\phi)\ {\rm d}\cos\theta \;,
\end{equation}
The longitudinal average:
\begin{equation}
\langle h \rangle_\phi \equiv \frac{1}{2\pi}\int_{-\pi}^{\pi}h(\theta,\phi)\ {\rm d}\phi \;,
\end{equation}
The spherical average: 
\begin{equation}
\langle h\rangle_s \equiv \frac{1}{4\pi}\int_{-1}^1 \int_{-\pi}^{\pi} h(\theta,\phi)\ {\rm d}\cos\theta  {\rm d}\phi \;,
\end{equation}
The northern hemispheric average:
\begin{equation}
\langle h\rangle_+ \equiv \frac{1}{2\pi}\int_{0}^1 \int_{-\pi}^{\pi} h(\theta,\phi)\ {\rm d}\cos\theta  {\rm d}\phi \;.
\end{equation}
The time-average of each spatial mean is denoted by additional angular brackets, such as $\langle \langle h \rangle_\theta \rangle$.  

\subsection{Properties of Convective Motion}
Figure~4 shows, in the Mollweide projection, the distribution of the radial velocity when $t=330\tau_c$ on spherical surfaces at different depths 
for two models. Panels~(a)--(c) correspond to the depths $r=0.95R$, $0.85R$ and $0.72R$ for Model~A, and panels~(d)--(f) are those 
for Model~B. The orange and blue tones depict upflow and downflow velocities. At the upper ($r=0.95R$) and mid ($r=0.85R$) convection zones, 
the convective motion is characterized by upflow dominant cells surrounded by networks of narrow downflow lanes for both models. The higher 
the latitude, the smaller the convective cell prevails. Elongated columnar convective cells aligned with the rotation axis appear near the equator. 
These are the typical features observed in rotating stratified convection (e.g., \citealt{spruit+90,miesch+00,brummell+02,brun+04}).  
In panel~(c), we find that the downflow lanes persist the plume-like coherent structure even just above the bottom of the unstable layer 
($r=0.72R$). The downflow plumes then penetrate into the underlying stable layer. 

The radial profile of the mean radial velocity $\langle\langle v_r^2 \rangle_{\rm s}\rangle^{1/2}$ is shown in Figure~5. The red-solid 
and blue-dashed curves correspond to Models~A and~B, respectively. The time average spans in the range of $300\tau_c \le t \le 400 \tau_c$. 
The mean radial velocity has a peak at the mid convection zone ($r\sim 0.8R$) for both models. The convective motion is the most active there.  
While the radial flow is restrained by the boundary placed on the bottom of the convection zone in Model~B, it can penetrate into the underlying 
stable layer in Model~A. As a result of the penetrative convection, mean zonal and meridional flows are driven by the Reynolds and Maxwell 
stresses in the stable layer. That will be described in the followings. 
\subsection{Structures of Mean Flow}
In Figures~6(a) and (b), time-averaged mean angular velocity, defined by $\langle \langle \Omega \rangle_\phi \rangle = 
\langle\langle v_{\phi} \rangle_\phi \rangle /(r \sin\theta) + \Omega_0 $, is shown for the models~A and~B, respectively. 
The time average spans in the range of $300\tau_c \le t \le 400 \tau_c$. The normalization unit is the initial angular velocity 
$\Omega_0$. 

The differential rotations in both models have basically solar-like profiles with the equatorial acceleration. However, both exhibit more cylindrical 
alignment than the solar rotation profile characterized by the conical iso-rotation surface. The system is dominated by the Taylor-Proudman 
balance in both models (e.g., \citealt{pedlosky87}). The angular velocity contrast $\Delta \Omega$ between equator and pole is about 
$18$\% in Model~A and $16$\% in Model~B. These are slightly smaller than that obtained by the helioseismology ($\sim 20$\%). 
More remarkably, a radial gradient of the angular velocity is developed in the stably stratified layer around latitudes $\pm 40^\circ$. 
This structure is reminiscent of the solar tachocline despite the radial shear layer is broad compared to the observed one 
(\citealt{spiegel+92,charbonneau+99,miesch05,hughes+07}). The rotation profile of Model~A is reasonably similar with that of the 
Sun deduced from helioseismology (\citealt{thompson+03}). 

The spontaneous formation of the tachocline-like shear layer below the convective envelope was reported in the hydrodynamic simulation 
of the solar penetrative convection performed by \citet{brun+11}. Our results suggest that the tachocline-like shear layer is a natural outcome of the
presence of the stable layer even in the MHD convection system. We discuss more about the differential rotation profile established in Model~A in \S~4.1. 

Shown in Figures~6(c) and (d) are time-averaged mean meridional flows for the models~A and~B. The color contour depicts the meridional flow velocity, 
defined by $\langle\langle v_m \rangle_\phi \rangle = [\langle\langle v_r \rangle_\phi \rangle^2 + \langle\langle v_\theta \rangle_\phi \rangle^2]^{1/2}$, with a maximum $\sim 0.1v_{\rm rms}$. Overplotted are streaklines with a length proportional to the flow speed. 
The circulation flow is primarily counter-clockwise in the bulk of the convection zone in the northern hemisphere, that is, the poleward in the 
upper convection zone and the equatorward in the bottom convection zone in both models. There is however a clear difference in the circulation 
pattern between the two models. While a large single-cell is formed in Model~B,  Model~A shows a double-cell pattern with a strong inward/outward 
flow at the low/mid latitudes. An intriguing finding is that the equatorward component penetrates into the underlying stable layer when the  
radial gradient of the angular velocity resides (see Figure~6(a)). This suggests that the penetrative transport of magnetic flux by the meridional 
flow might play a role in magnetic dynamo in our model.  

The meridional flow takes a role in transporting angular momentum and magnetic flux in the Sun. However, the circulation pattern, velocity, 
and their time variations are still controversial as compared with the mean angular velocity profile which is well-confirmed by the helioseismic 
measurement. This is because the meridional circulation is much weaker than the differential rotation. Although the global circulation consisting 
of two cells is implied by the helioseismic inversion (see \citealt{mitra-kraev+07}) and is also obtained by numerical simulation 
(e.g., \citealt{miesch+06}), it is not agreed in general (\citealt{hathaway12,schad+12}). 

The mean-field theory of the angular momentum transport predicts a single-cell circulation (c.g., \citealt{rudiger89}). Despite the kinematic 
flux-transport dynamo model is constructed based on the single-cell circulation (\citealt{dikpati+99,charbonneau05}), 
the influence of the circulation pattern on the magnetic activity is still a matter of debate (\citealt{pipin+13} and references therein). 
In order for more detailed discussion about the meridional flow, we should improve our simulation model in such a way to achieve 
smaller-scale subsurface convection ranging from granulation to super-granulation as probed by local helioseismology 
(\citealt{gizon+05,rieutord+10}). 
\subsection{Structures of Magnetic Field}
\begin{figure}[tpb]
\begin{center}
\scalebox{1.1}{{\includegraphics{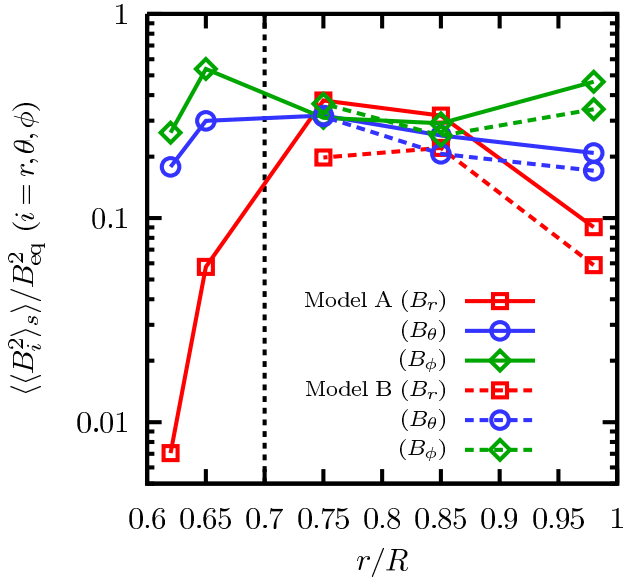}}} 
\caption{Time and surface average of the magnetic energy density  as a function of radius. The sampled radii are
$r/R=0.62,0.65,0.75,0.85, 0.98$ for Model~A, and $r/R=0.75, 0.85, 0.98$ for Model~B. The broken solid lines with red-squares, 
blue-circles and green-diamonds denote the contributions from radial, latitudinal and azimuthal components of the magnetic field 
for Model~A. The broken dashed lines with the same symbols denote those for Model~B. The time average spans in the range of 
$100\tau_c \le t \le 400 \tau_c$. The vertical axis is normalized by $B_{\rm eq} = 0.02$}.
\label{fig7}
\end{center}
\end{figure}
\begin{figure*}[tpb]
\begin{center}
\begin{tabular}{cc}
\scalebox{1.05}{{\includegraphics{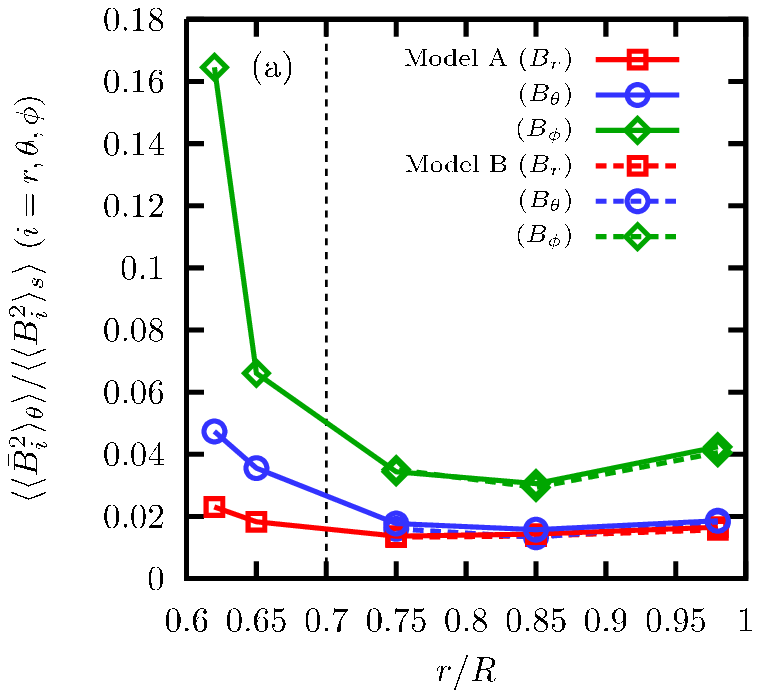}}} &
\scalebox{1.05}{{\includegraphics{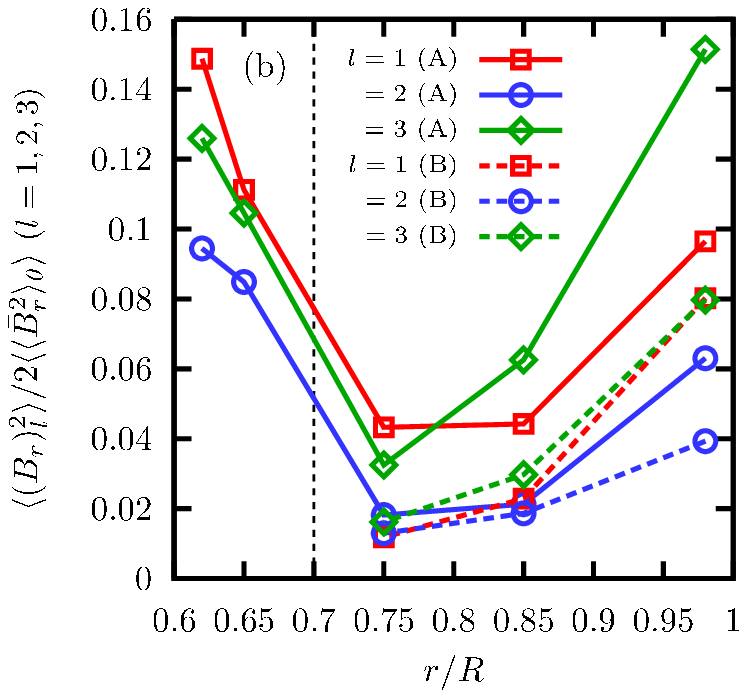}}} 
\end{tabular}
\caption{(a) profiles of $\langle \langle \bar{B}_i^2\rangle_\theta\rangle/\langle \langle B^2 \rangle_s \rangle $ for $i = r,\theta,\phi$.
(b) profiles of $\langle (\bar{B}_r)_l^2 \rangle /2\langle \langle \bar{B}_r^2 \rangle_\theta\rangle$ for $l = 1,2$ and $3$. The broken solid lines 
with red-squares, blue-circles and green-diamonds denote the radial, latitudinal and azimuthal components in panel~(a),  dipole ($l=1$), quadrupole 
($l=2$), and octupole ($l=3$) moments in panel~(b) for Model~A. The broken dashed lines with the same symbols denote those for Model~B. 
The time average spans in the range of $100\tau_c \le t \le 400 \tau_c$. }
\label{fig8}
\end{center}
\end{figure*}
\begin{figure*}[tpb]
\begin{center}
\scalebox{0.82}{{\includegraphics{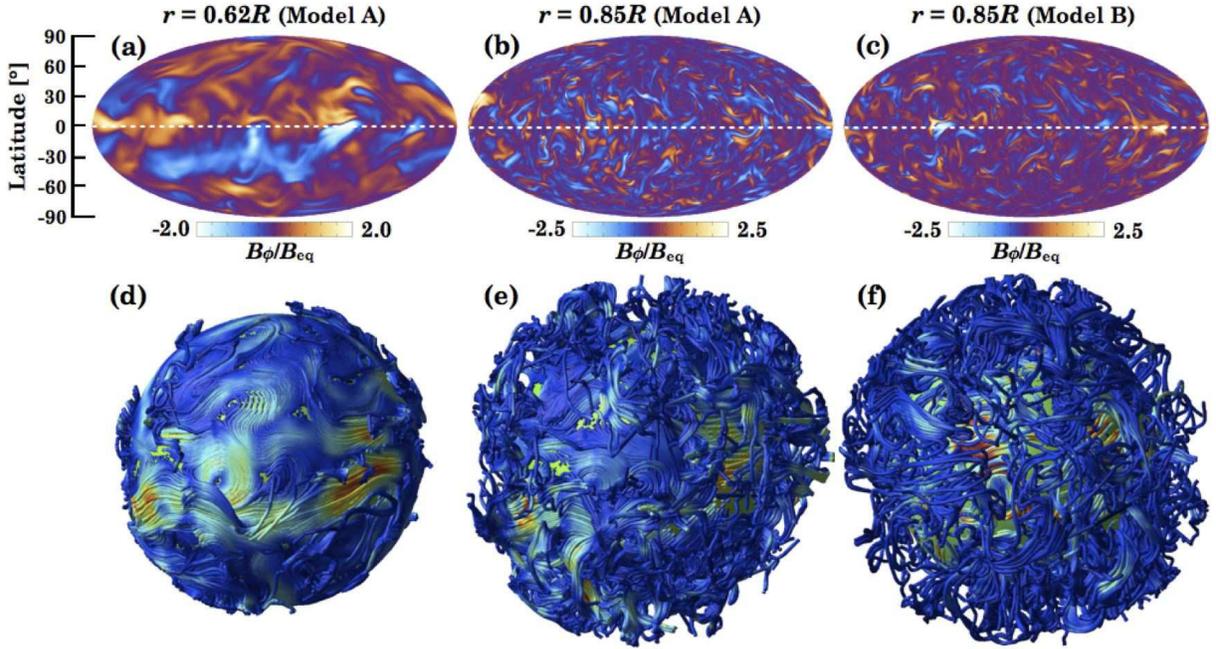}}} 
\caption{Snapshot of the azimuthal component of the magnetic field $B_{\phi}$ when $t=330\tau_c$ on a spherical surface at sampled  
radii (a) $r=0.62R$ and (b) $r=0.85R$ for Model~A, and (c) $r=0.85R$ for Model~B. The orange and blue tones depict positive 
and negative values of the $B_\phi$ component. The magnetic field lines at the time and position corresponding to those in 
the panels~(a)--(c) are visualized in the panels~(d)--(f), respectively}
\label{fig9}
\end{center}
\end{figure*}
\begin{figure*}
\begin{center}
\begin{tabular}{cc}
\scalebox{1.1}{{\includegraphics{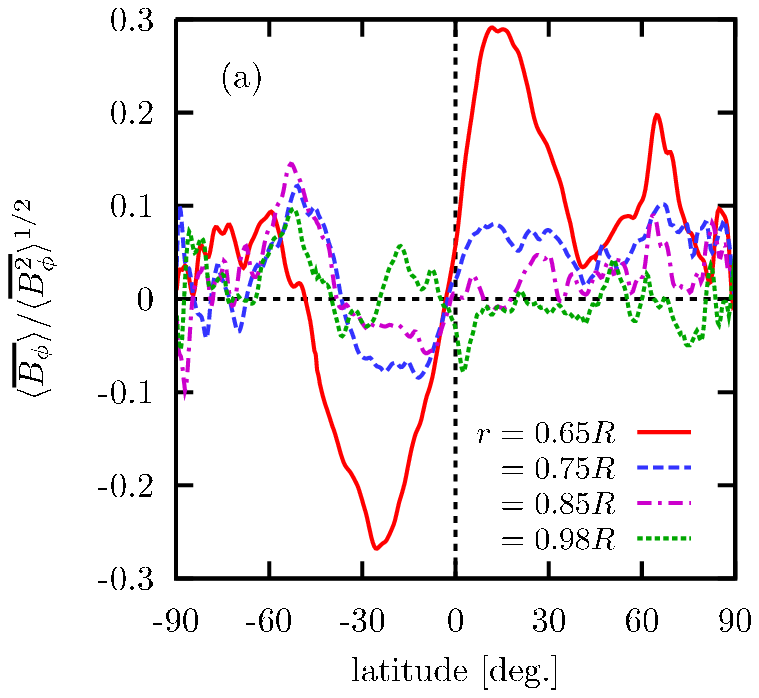}}} &
\scalebox{1.1}{{\includegraphics{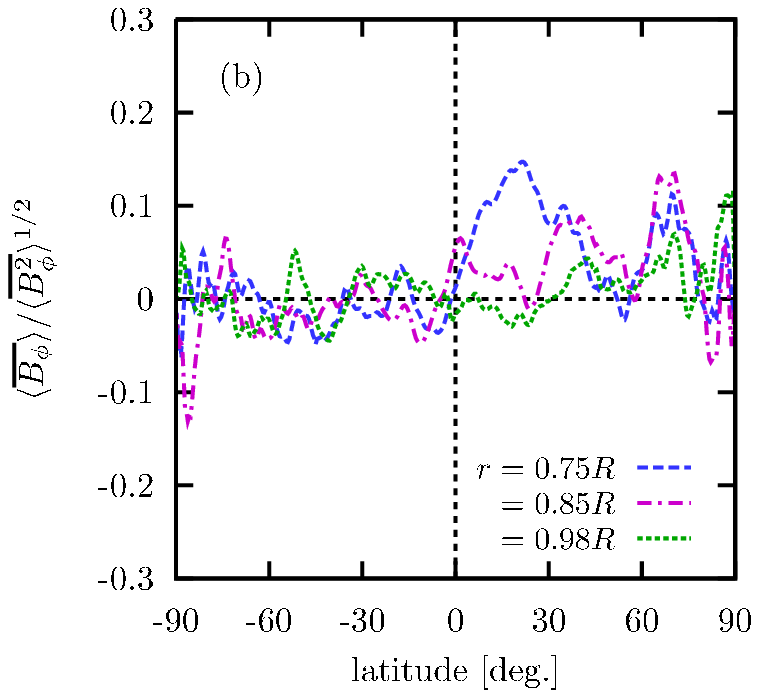}}} 
\end{tabular}
\caption{ Latitudinal profiles of $\langle \bar{B_\phi} \rangle  /\langle\overline{B_\phi^2} \rangle^{1/2}$ at the sampled radii for 
Models~A and~B. The red-solid, blue-dashed, purple-dash-dotted and green-dotted curves correspond to $r=0.65R$, $0.75R$, $0.85R$, 
and $0.98R$, respectively. The time average spans in the range of $240\tau_c \le t \le 340 \tau_c$ for Model~A or 
$310\tau_c \le t \le 360 \tau_c$ for Model~B. } 
\label{fig10}
\end{center}
\end{figure*}
\begin{figure}[tpb]
\begin{center}
\scalebox{0.43}{{\includegraphics{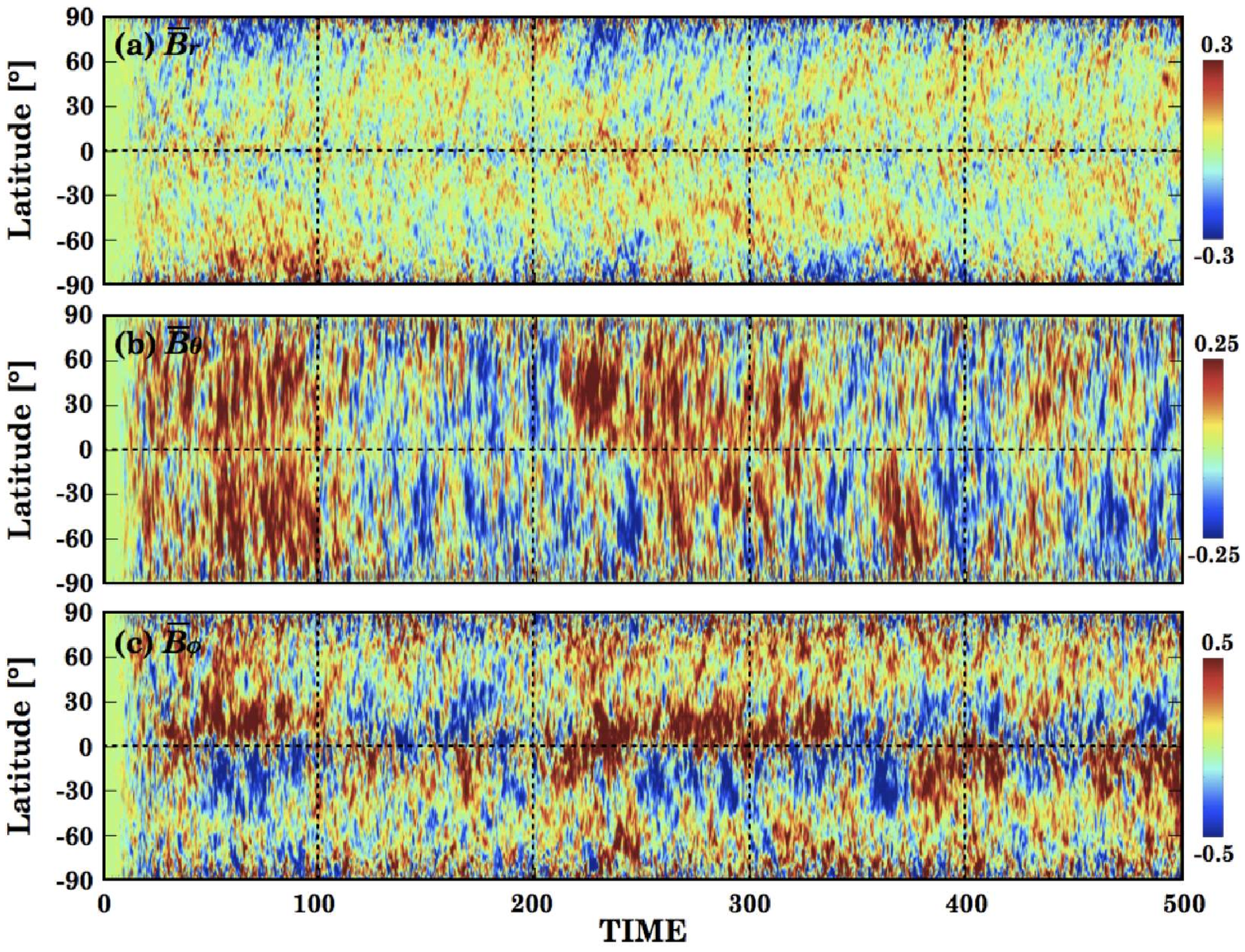}}} 
\caption{Azimuthally-averaged magnetic field as a function of time and latitude for Model A. The top, middle and bottom panels correspond to 
$\bar{B}_r$, $\bar{B}_\theta$, and $\bar{B}_\phi$ at the depth $r=0.65R$. The red and blue tones depict positive and negative values of 
each magnetic component. }
\label{fig11}
\end{center}
\end{figure}
\begin{figure}[tpb]
\begin{center}
\scalebox{0.43}{{\includegraphics{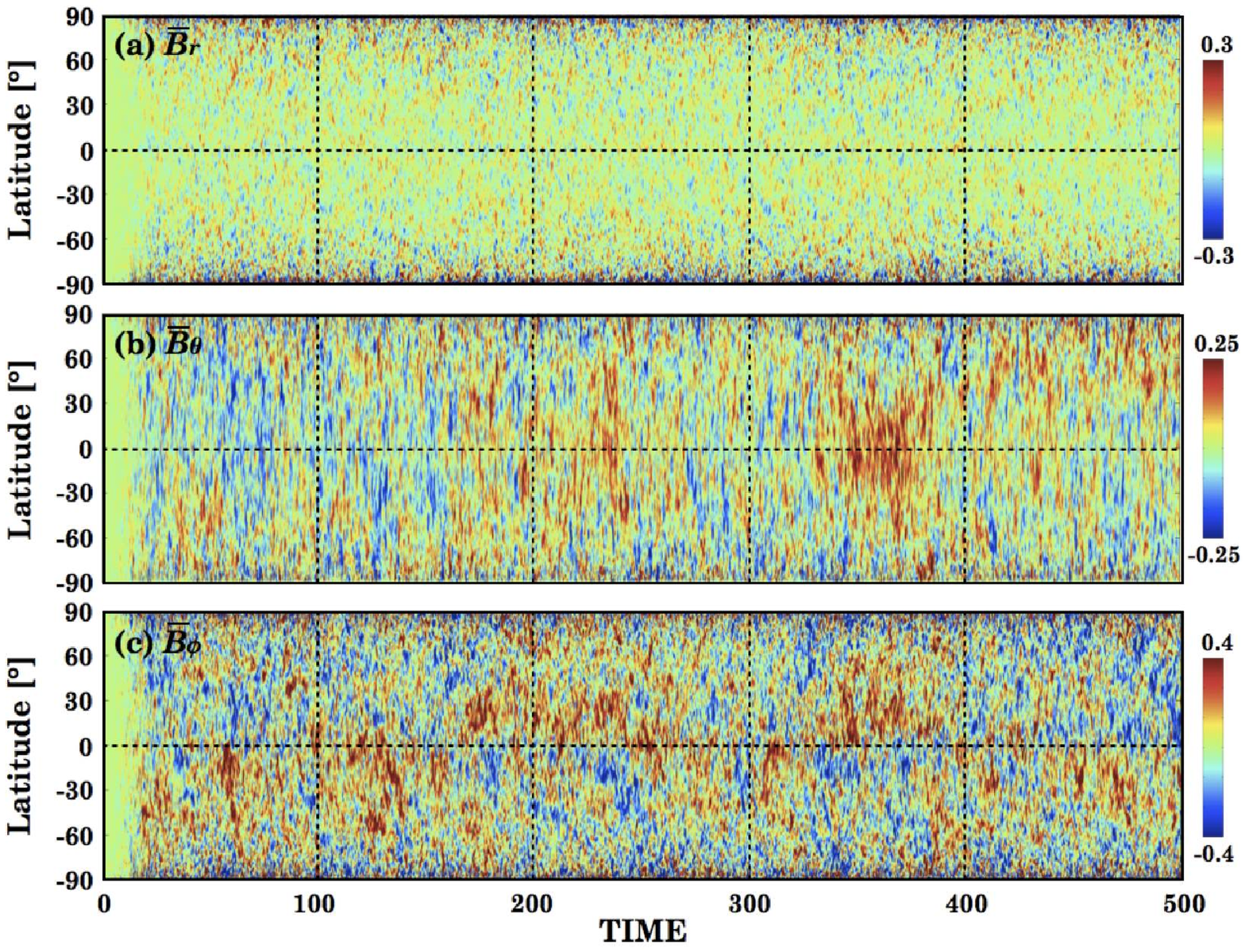}}} 
\caption{Azimuthally-averaged magnetic field as a function of time and latitude for Model B. The top, middle and bottom panels correspond to 
$\bar{B}_r$, $\bar{B}_\theta$, and $\bar{B}_\phi$ at the depth $r=0.72R$. The red and blue tones depict positive and negative values of 
each magnetic component. }
\label{fig12}
\end{center}
\end{figure}
As shown in Figure~3, the magnetic energy is amplified by the dynamo action and is saturated at a level of about $40$\% of the convective 
kinetic energy for both models after $t\simeq 30\tau_c $. The magnetic field is maintained longer enough than the magnetic diffusion 
time ($\sim 100\tau_c$). Although the volume-averaged magnetic energy is almost the same in the models~A and~B, there 
are remarkable differences in the spatial structure of the magnetic fields. 

The time and surface average of the magnetic energy density is presented as a function of radius in Figure~7. 
The broken solid lines with red-squares, blue-circles and green-diamonds denote the contributions from radial, latitudinal and azimuthal 
components of the magnetic field for Model~A. The broken dashed lines with the same symbols denote those for Model~B. The time average spans in the range of 
$100\tau_c \le t \le 400 \tau_c$. While the contributions of three magnetic components are almost  the same at the mid convection zone 
($r\simeq 0.85$) where the convective motion is vigorous, the azimuthal component becomes predominant in the region where the convective 
motion is less active for both models (see also Figure~5). Especially, in the stable layer of Model~A, most of the magnetic energy is stored 
as a form of the azimuthal field. 

To examine the magnetic structure in more detail, we divide the magnetic energy density into axi-symmetric part and asymmetric part (see Appendix)
\begin{equation}
\langle \bm{B}^2 \rangle_s = \langle \bar{B}_r^2 \rangle_\theta  + \langle \bar{B}_\theta^2 \rangle_\theta +  \langle \bar{B}_\phi^2 \rangle_\theta 
+ ({\rm asymmetric \ part}) \;,
\end{equation}
where we denote the axi-symmetric part of the magnetic field $B_i$ for $i={r,\theta,\phi}$, 
\begin{equation}
\bar{B_i} \equiv \langle B_i \rangle_\phi \;.
\end{equation}

To elucidate the relative strengths of axi-symmetric components, we plot the profiles of 
$\langle \langle \bar{B}_i^2\rangle_\theta\rangle/\langle \langle B^2 \rangle_s \rangle$ in Figure 8(a). 
The broken solid lines with red-squares, blue-circles and green-diamonds denote the radial, latitudinal and azimuthal components for Model~A. 
The broken dashed lines with the same symbols denote those for Model~B. The time average spans in the range of $100\tau_c \le t \le 400 \tau_c$. 
Among the three axi-symmetric components, $\bar{B}_\phi$ is dominant. The tendency of $\bar{B}_\phi$--dominance is apparent not only in the stable 
layer, but also in the convection zone both in models A and B. The relative strength of the axi-symmetric component increases with the depth and reaches 
the maximum at around the bottom stable zone. Figures 7 and 8(a) suggest that the axi-symmetric component is built up rather in the convectively 
calm layer although the magnetic energy is amplified preferentially in the region where the vigorous convective motion exits. 

We then analyze the latitudinal moments of the axi-symmetric field $\bar{\bm{B}}  $. We focus on $\bar{B}_{r}$ since this 
component reflects purely the poloidal field, while $\bar{B}_\theta$ and $\bar{B}_\phi$ are mixture of the toroidal and poloidal fields. 
From the Peseval's equation (see Appendix), 
\begin{equation}
\langle \bar{B}_r^2 \rangle_\theta = \frac{1}{2} \sum_{l=1}(\bar{B}_r)_l^2 \;,
\end{equation}
where
\begin{equation}
(\bar{B}_r)_l = \int_{-1}^{1} \bar{B}_r P_l^*(\cos\theta)\ {\rm d}\cos\theta \;.
\end{equation}
Here $P_l^*$ are normalized Legendre polynomials. Figure 8(b) shows profiles of 
$\langle (\bar{B}_r)_l^2 \rangle /2\langle \langle \bar{B}_r^2 \rangle_\theta\rangle$ for $l = 1,2$ and $3$. The broken solid lines with red-squares, 
blue-circles and green-diamonds denote  dipole ($l=1$), quadrupole ($l=2$), and  octupole ($l=3$) moments for Model A. The broken dashed lines with the 
same symbols denote those for Model B. The time average spans in the range of $100\tau_c \le t \le 400 \tau_c$. There is not much difference among amplitudes of dipole, quadrupole and octupole moments at all the depth. Nevertheless, it would be worth noting that Model~B has a octupole dominance 
in almost the whole domain. In the case of Model~A, the dipole gradually becomes dominant with the depth. It is predominant in the bottom 
convection zone and the stable zone although the upper and mid convection zones are dominated by higher multipoles like as Model~B. The stably 
stratified layer below the convective envelope promotes dipole solution as indicated by \citet{miesch+09}.

The similarity and difference of the magnetic structure between two models are the most obvious on the azimuthal component of the magnetic 
field. A snapshot of the azimuthal component of the magnetic field at $t=330\tau_c$ is presented in Figure~9 on a spherical surface at 
(a) $r=0.65R$ and (b) $r=0.85R$ for Model~A, and (c) $r=0.85R$ for Model~B. The orange and blue tones depict positive and negative 
values of the $B_\phi$ component normalized by $B_{\rm eq}$. The magnetic field lines at the time corresponding to those in the 
panels~(a)--(c) are visualized in Figures~9(d)--(f), respectively. As expected from Figures~7 and 8, the convective envelope is dominated by disordered 
tangled magnetic field lines with a myriad of localized small-scale structures in both models. These incoherent magnetic fields are strongly 
influenced by vigorous convective motions and thus are highly intermittent. The horizontal converging flows sweep magnetic fields into downflow 
lanes and intensify them locally to the super-equipartition strength as was observed in existing convective dynamo simulations 
(e.g., \citealt{brandenburg+96,cattaneo+03,brun+04}). 

In the underlying stable layer of Model~A, a strong large-scale azimuthal component of magnetic field is built up around the equator, and resides 
there for long time intervals. This well-organized magnetic component is roughly antisymmetric about the equatorial plane and has a maximum 
strength of an order of $B_{\rm eq}$. The large-scale component is organized in the stable zone where the radial angular velocity gradient 
resides (see Figure~6(a)). This would be an important evidence of a connection between the deep-seated large-scale magnetic component and 
the tachocline-like shear layer that is spontaneously developed in the model with the stable layer. 

The latitudinal profiles of $\langle \bar{B_\phi} \rangle  /\langle\overline{B_\phi^2} \rangle^{1/2}$ are shown at the 
sampled radii in Figures~10(a) and~(b) for Models~A and~B. The red-solid, blue-dashed, purple-dash-dotted and green-dotted 
curves correspond to $r=0.65R$, $0.75R$, $0.85R$, and $0.98R$, respectively. The time average spans in the range of 
$240\tau_c \le t \le 340 \tau_c$ for Model~A or $310\tau_c \le t \le 360 \tau_c$ for Model~B. As shown in Figures~8(a) and~9, the strong mean 
azimuthal component with the antisymmetric profile is built up around the equatorial plane in the stable layer of Model~A. It reaches maximum strength 
at around the latitude $\pm 25^{\circ}$. While the antisymmetric property of the mean-field is found not only in the stable layer but also in the 
convective envelope, the amplitude of the mean-field component is much smaller in the convection zone than in the stable zone.  In comparison with 
the Model~A, the antisymmetric property of the mean azimuthal field is weaker in the Model~B. 

Overall magnetic structures simulated in our models indicate that the stably stratified layer is an important building block to organize large-scale 
magnetic components and support numerical studies of \citet{browning+06} and \citet{ghizaru+10}. Despite the magnetic energy is amplified by the 
vigorous convective motion in the mid convection zone, the strong mean magnetic component is preferentially organized in the region where the 
convective motion is less vigorous. This suggests that the downward pumping process of the magnetic flux is of great importance in the solar 
dynamo mechanism (\citealt{tobias+01,tobias+08,barker+12}). The implementation of more realistic convective penetration and downward 
pumping processes into the numerical modeling might come the first to reproduce the solar dynamo.
\subsection{Cyclic Property of Magnetic Fields}
\begin{figure*}[tbp]
\begin{center}
\begin{tabular}{cc}
\scalebox{0.41}{{\includegraphics{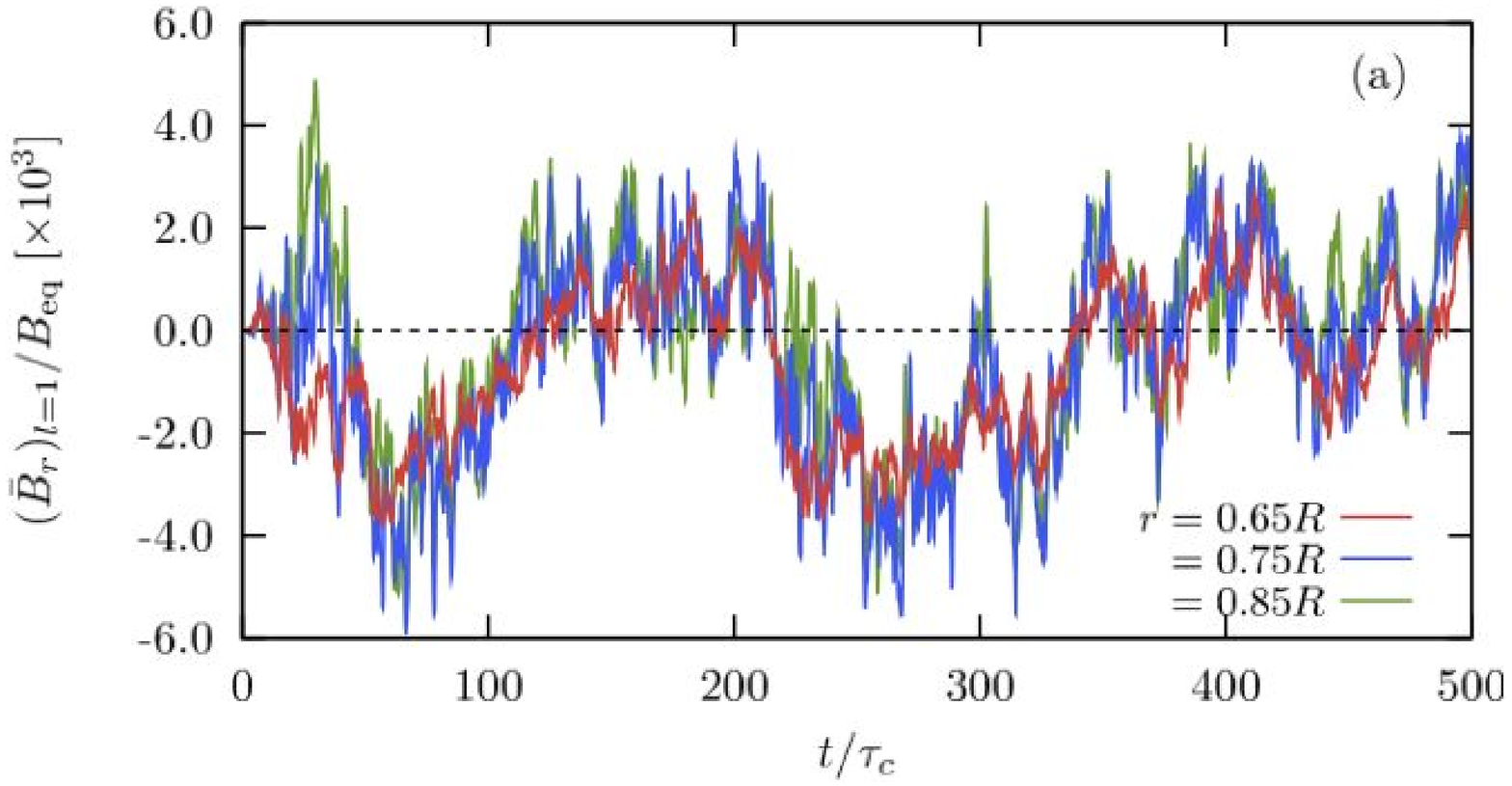}}} &
\scalebox{0.41}{{\includegraphics{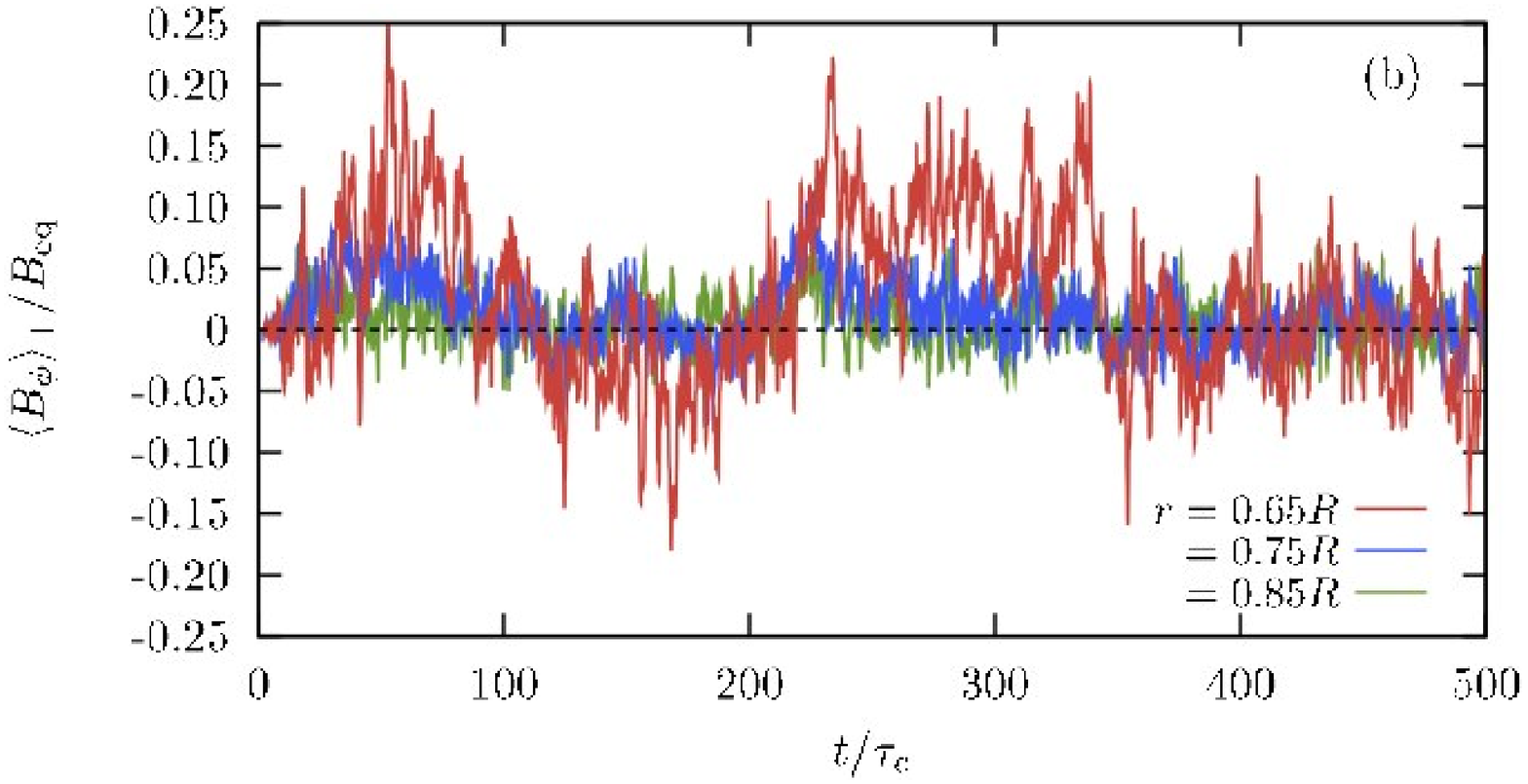}}} \\
\scalebox{0.41}{{\includegraphics{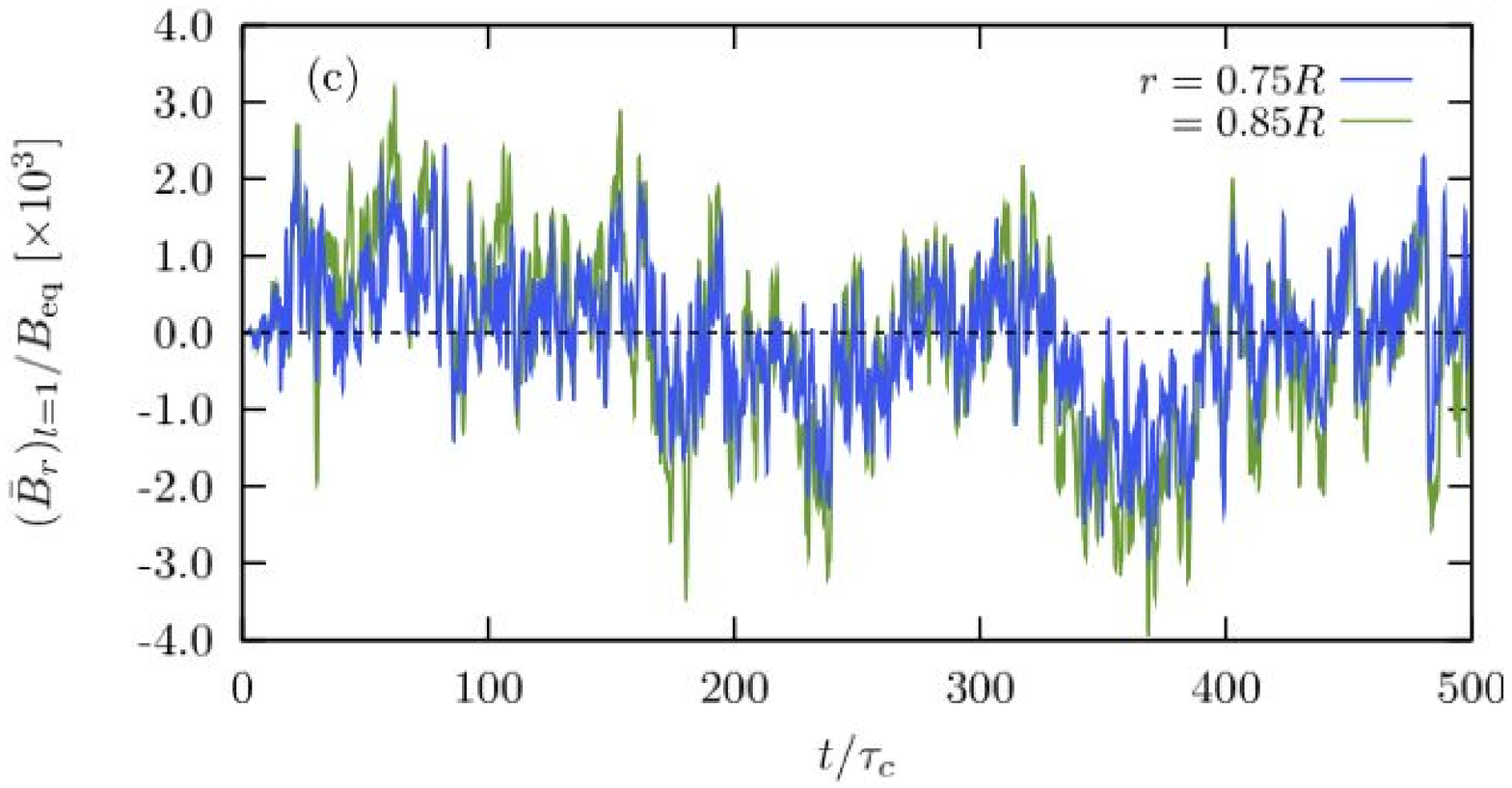}}}  &
\scalebox{0.41}{{\includegraphics{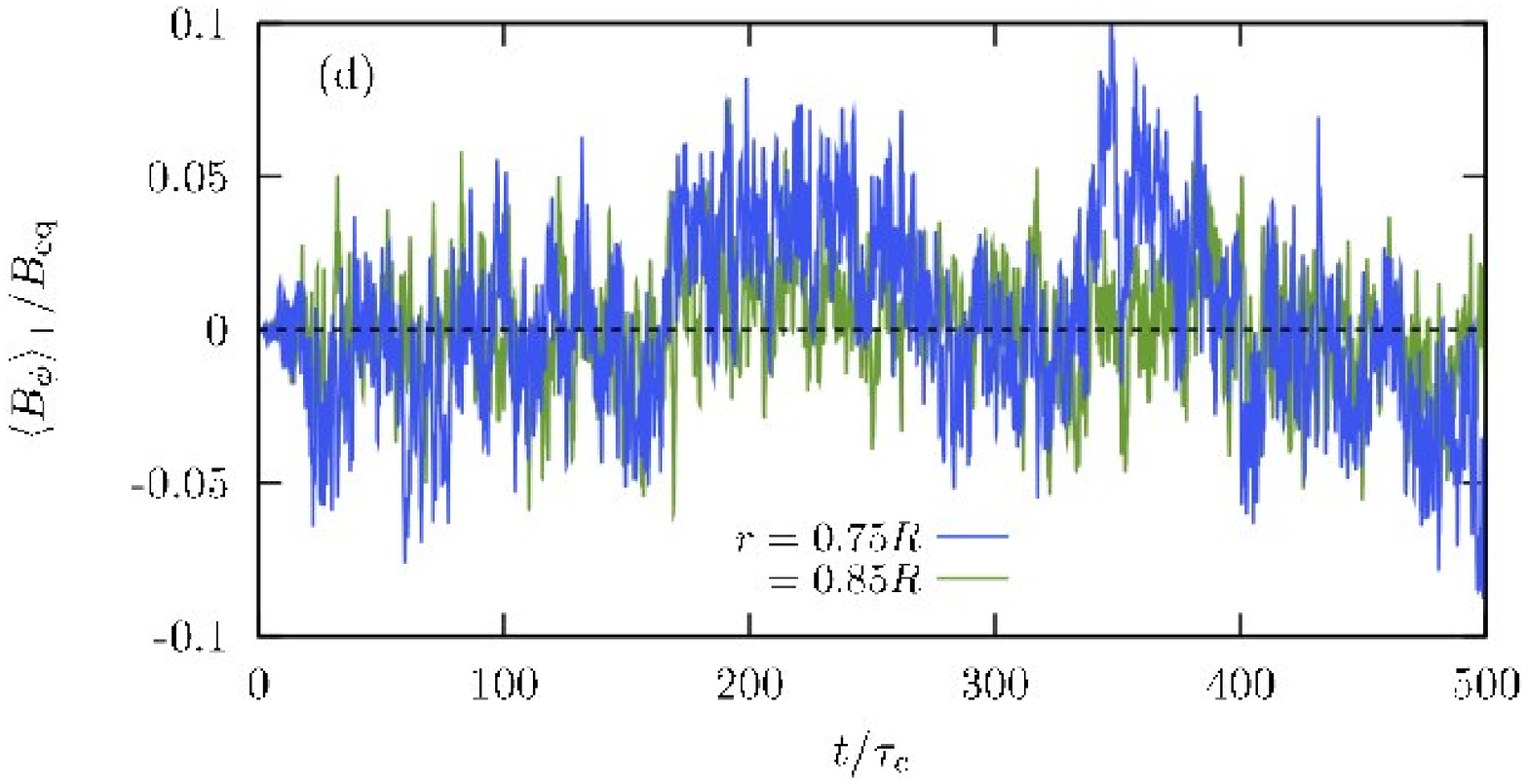}}} 
\end{tabular}
\caption{Temporal evolutions of (a) the dipole moment $(\bar{B}_r)_{l=1}$ and (b) the northern hemispheric average of the azimuthal field 
$\langle B_\phi\rangle_+$. Panels~(a) and (b) correspond to $(\bar{B}_r)_{l=1}$ and  $\langle B_{\phi}\rangle_+$ for Model~A. 
Panels~(c) and (d) are those for Model~B. The red, blue, and green curves correspond to the depths $r=0.65R$, $0.75R$, and $0.85R$, 
respectively. } 
\label{fig13}
\end{center}
\end{figure*}
One of the most interesting findings in our simulation is that the large-scale magnetic fields show polarity reversals.  Figure~11 gives an 
azimuthally-averaged magnetic field as a function of time and latitude for Model~A. The panels~(a), (b) and (c) represent $\bar{B}_r$, 
$\bar{B}_\theta$, and $\bar{B}_\phi$ at  $r=0.65R$. The large-scale $\bar{B}_\phi$ with antisymmetric parity persists over a relatively 
long period despite strong stochastic disturbances due to penetrative convective motions. The $\bar{B}_\phi$ component changes the sign 
for at least three times, at about $t=100\tau_c$, $t=210\tau_c$, and $t=350\tau_c$ [panel (c)]. 

As for the poloidal component of the magnetic field, it shows the dipole dominance in the stable layer as indicated in Figure~8(b). During 
$30$--$100\tau_c$ when strong $\bar{B}_\phi$ component with positive polarity dominates in the northern hemisphere, negative $\bar{B}_r$ 
and positive $\bar{B}_\theta$ are observed [panels~(a) and~(b)]. While the $\bar{B}_\theta$ component has the same sign in both the 
hemispheres, the $\bar{B}_r$ component has opposite polarity in the two hemispheres. When the $\bar{B}_\phi$ reversal takes place, 
the other two components $\bar{B}_r$ and $\bar{B}_\theta$ also change the sign. See, for example, at about $t=100\tau_c$ in Figure~11. 

The time-latitude diagram of an azimuthally-averaged magnetic field for Model~B is shown in Figure~12. The panels~(a),~(b) and~(c) represent 
$\bar{B}_r$, $\bar{B}_\theta$, and $\bar{B}_\phi$ at  $r=0.75R$. The mean-field component shows a week polarity preference 
and polarity reversals in time even in the model without the stable layer. While the $\bar{B}_\phi$ component has an antisymmetric profile 
with respect to the equator, the $\bar{B}_\theta$ component has the same sign in both the hemispheres as well as the Model~A. 
However, the amplitude, coherency and dipole dominance of the mean magnetic component are much weaker in the model~B 
compared with those in the model~A. These are consistent with Figures~8--10. 

As the indicators of the depth-dependency of the polarity reversal, we show the temporal evolution of the dipole moment $(\bar{B}_r)_{l=1}$ defined 
by equation (18) and the northern hemispheric average of the azimuthal field $\langle B_\phi\rangle_+$ in Figure~13. Panels~(a) and (b) 
correspond to $(\bar{B}_r)_{l=1}$ and  $\langle B_{\phi}\rangle_+$ for Model~A. Panels~(c) and (d) are those for Model~B. The red, blue, and green 
curves correspond to the depths $r=0.65R$, $0.75R$, and $0.85R$, respectively. 
The polarity reversal takes place not only in the underlying stable layer but also in the convective envelope for Model A. 
It is remarkable that there is a clear phase synchronization in the polarity reversals at different depths. Even in the case of the Mode B, 
the polarity reversal of the mean-field component is noticeable although the short-term variability is superimposed onto the 
global long-term modulation. The cycle period is about $100\tau_c$ for both models.  
This indicates that the polarity reversal of the mean-field component is a global phenomenon that takes place synchronously 
throughout the system regardless the presence of the stable layer. 

\section{Discussion}
\subsection{Force Balance in Differential Rotation}
\begin{figure*}[tbp]
\begin{center}
\begin{tabular}{cc}
\scalebox{1.1}{{\includegraphics{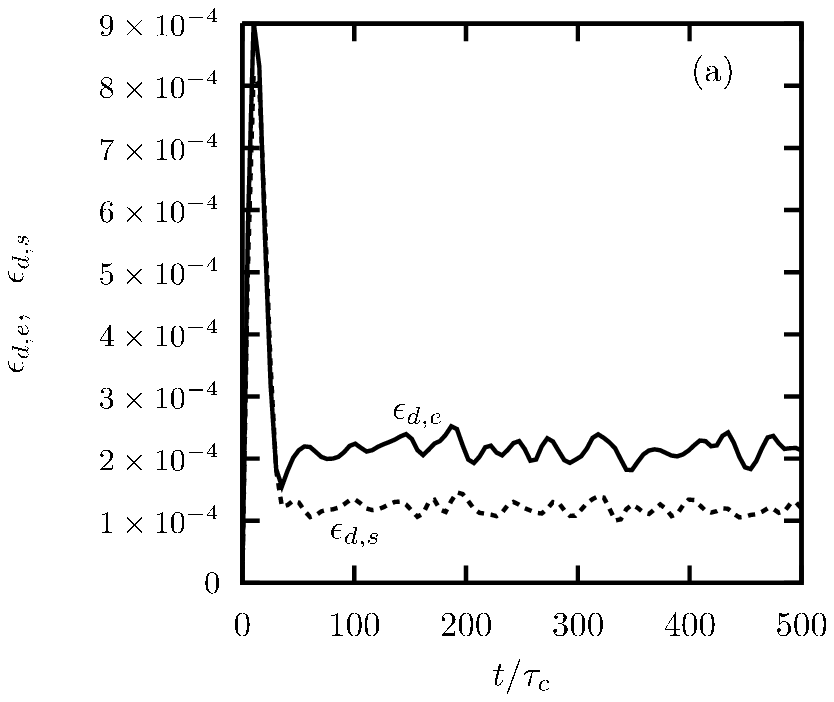}}} &
\scalebox{1.1}{{\includegraphics{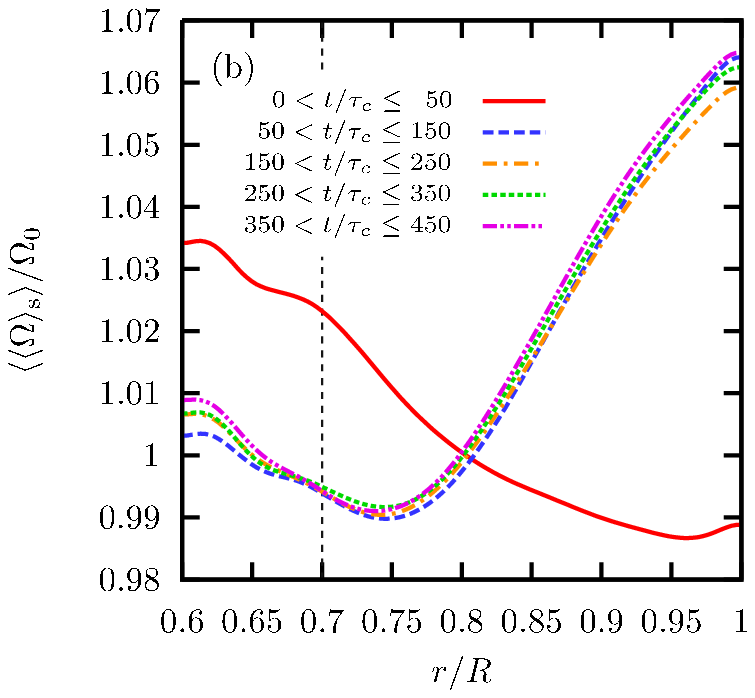}}} 
\end{tabular}
\caption{(a) Temporal evolution of the kinetic energy of differential rotation averaged over the entire volume ($\equiv \epsilon_{d,e}$) and 
over the radiative zone ($\equiv \epsilon_{d,s}$). (b) Radial profile of the mean angular velocity averaged over a given time span 
$\langle\langle \Omega \rangle_{\rm s}\rangle$ for Model~A. The different line types correspond to different time spans. Note that the 
surface average is here taken over the range of $60^\circ \le \theta \le 120^\circ$ (around the equator). } 
\label{fig14}
\end{center}
\end{figure*}
\begin{figure*}[tbp]
\begin{center}
\begin{tabular}{cc}
\scalebox{1.1}{{\includegraphics{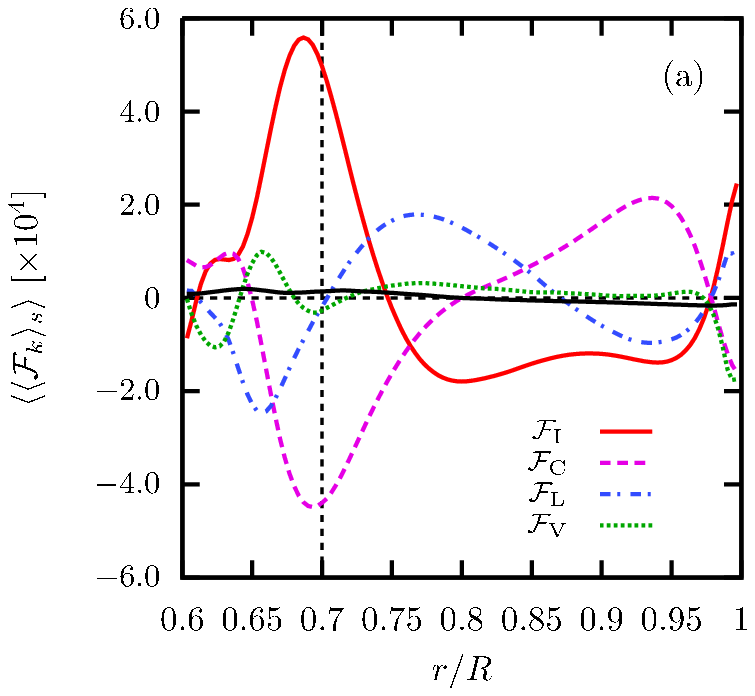}}} &
\scalebox{1.1}{{\includegraphics{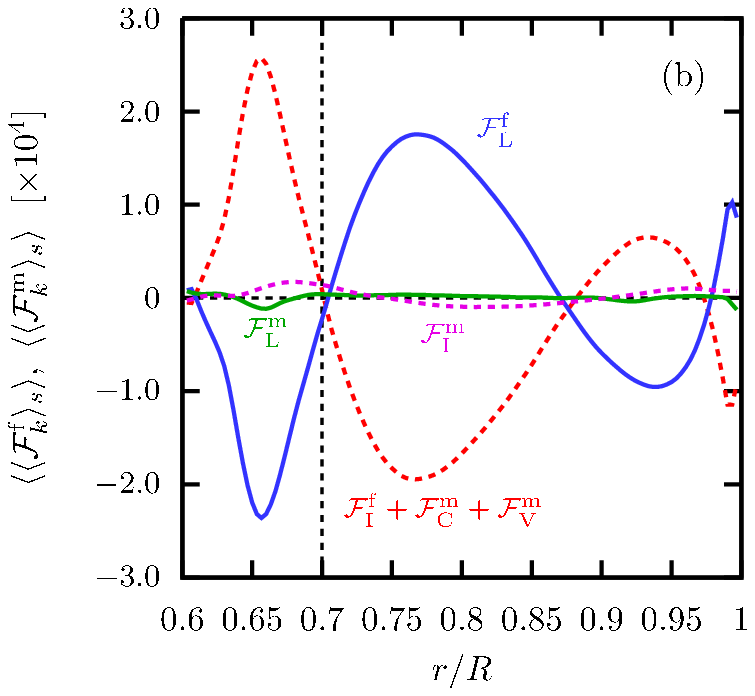}}} 
\end{tabular}
\caption{(a) Time and surface averages of various force terms as function of radius. The solid red, dashed magenta, dash-dotted blue, 
dotted green curves correspond to the radial profiles of the inertia, Coriolis, Lorentz and viscous forces, respectively. The solid black curve 
denotes the sum of the four azimuthal forces. (b) The radial profiles of $\langle\langle \mathcal{F}_{\rm I}^{\rm f}\rangle_s\rangle+\langle\langle \mathcal{F}_{\rm C}^{\rm m}\rangle_s\rangle+\langle\langle\mathcal{F}_{\rm V}^{\rm m}\rangle_s\rangle$ and $\langle\langle \mathcal{F}_{\rm I}^{\rm m} \rangle_s\rangle$ by the red dashed and magenta dashed curves. The blue and green solid curves denote the radial profiles of 
$\langle\langle\mathcal{F}_{\rm L}^{\rm f} \rangle_s\rangle$ and $\langle\langle\mathcal{F}_{\rm L}^{\rm m} \rangle_s\rangle$. } 
\label{fig15}
\end{center}
\end{figure*}
The helioseismic measurements suggest that the tachocline thickness is less than $4$\% of the solar radius (e.g., \citealt{elliott+99,
charbonneau+99,basu+01}). The presence of such a thin transition layer leads to the tachocline confinement problem (\citealt{spiegel+92}). 
The mechanism that inhibits the differential rotation in the convection zone to spread into the deeper radiative interior is still an open problem, 
though several theoretical models have been proposed (\citealt{rudiger+97,gough98,rogers11,brun+11}). 

In conjunction with the tachocline confinement problem, we examine the time evolution of the differential rotation established in our simulation 
model with the stably stratified layer (Model A). We show, in Figure~14~(a), the temporal evolution of the kinetic energy of differential rotation 
averaged over the entire volume ($\equiv \epsilon_{d,e}$) and the volume of the stable 
zone ($\equiv \epsilon_{d,s}$ ) defined by 
\begin{eqnarray}
\epsilon_{d,e} & = & \int_{r\le1.0R} \left[ \frac{1}{2}\rho v_\phi^2 \right] dV \Big / \int_{r\le1.0R} dV \;, \nonumber \\
\epsilon_{d,s} & = & \int_{r\le0.7R} \left[ \frac{1}{2}\rho v_\phi^2 \right] dV \Big / \int_{r\le0.7R} dV \;. 
\end{eqnarray} 
After the initial transitional stage $t \lesssim 50\tau_c$, the kinetic energy of the differential rotation is settled into an approximately 
constant value both in the entire domain and in the stable zone. In our simulation, the viscous spreading of the tachocline-like shear layer 
operates on a timescale of $\tau_{\rm vis} \sim 50\tau_{c}$. This dominates over the radiative spreading controlled by the Eddington-Sweet 
timescale. Since the duration of the simulation is about $10\tau_{\rm vis}$, the differential rotation has achieved the equilibrated profile. 

The statistical stationarity of the differential rotation profile is confirmed in Figure 14~(b), which shows the radial profile of the mean angular 
velocity averaged over a given time span, $\langle\langle \Omega \rangle_{\rm s}\rangle$, for Model~A. The different lines correspond 
to different time spans. Here the surface average is taken over the range of $60^\circ \le \theta \le 120^\circ$ (around the equator). 
After the initial evolutionary stage ($t \lesssim 50\tau_c$), the differential rotation attains a stationary profile in which outer shell is rotating faster. 
The differential rotation in the convection zone does not spread downward into the stable layer as time passes in spite of the shear on the interface 
of the convection and stable layers. 

The formation of the solar-like rotation profile is associated with the development of the magnetic field. As seen in the mean rotation profile during 
$0 \le t \lesssim 50\tau_c$, the inner shell rotates faster than the outer shell at the early dynamo kinematic stage. When the magnetic field is 
sufficiently amplified, it begins to affect the convective motion. The rotation profile changes to the opposite state in which the outer shell is rotating faster. 
This implies that the dynamo-generated magnetic field plays an important role in establishing the solar-like differential rotation. 

To elucidate the azimuthal force balance maintaining the differential rotation, we will consider the azimuthal component of momentum equation. 
The right-hand side of the equation is divided into four force terms: 
\begin{eqnarray}
\mathcal{F}_{\rm I} (\rho, \bm{v}, \bm{B}) &\equiv& [-{\rm div}(\rho \bm{v}\bm{v})]_\phi \;, \nonumber \\
\mathcal{F}_{\rm C} (\rho, \bm{v}, \bm{B}) &\equiv& [2 \rho \bm{v} \times \bm{\Omega}]_\phi \;, \nonumber \\
\mathcal{F}_{\rm L} (\rho, \bm{v}, \bm{B})&\equiv& [(\nabla \times \bm{B} ) \times \bm{B}]_\phi \;, \nonumber \\
\mathcal{F}_{\rm V} (\rho, \bm{v}, \bm{B})&\equiv& \left[ \mu ( \nabla^2 \bm{v} + \nabla (\nabla\cdot \bm{v})/3 ) \right]_\phi \;,
\end{eqnarray}
where $\mathcal{F}_{\rm I} $ is the inertia force, $\mathcal{F}_{\rm C} $ is the Coriolis force, $\mathcal{F}_{\rm L} $ is the Lorentz force, 
and $\mathcal{F}_{\rm V} $ is the viscous force. Note that azimuthal pressure gradient force does not contribute to the mean azimuthal force balance. 
These four force terms should cancel out each other for retaining the statistical equilibrium. 

The time and surface average of the each force term is demonstrated as a function of radius in Figure 15~(a). The solid red, dashed magenta, 
dash-dotted blue, dotted green curves correspond to the radial profiles of the inertia, Coriolis, Lorentz and viscous forces, respectively. 
The solid black curve denotes the sum of the four azimuthal forces. The time average is taken over $280\tau_c \le t \le 300\tau_c$ with $200$ snapshot 
data. The azimuthal force balance is mainly dominated by the inertia, Coriolis and Lorentz forces. In the convection zone, the negative inertia force 
balances with the sum of the positive Coriolis and Lorentz forces. In contrast to that, the positive inertia force is compensated by the sum of the negative 
Coriolis and Lorentz forces in the stable zone. The positive peak of $\mathcal{F}_{\rm I}$ below the interface between the convection zone and stable zone 
indicates the angular momentum transport by the penetrative convection. The viscous force makes a minor contribution to the azimuthal force balance 
except the surface layer and the bottom of the stable zone.  The net azimuthal force represented by solid black curve is nearly zero, confirming the statistical 
equilibrium of the azimuthal flow not only in the convection zone, but also in the stable zone. 

To examine the azimuthal force balance in more detail, we divide each force term into the contributions of the axisymmetric mean components 
given by $\mathcal{F}_{k}^{\rm m} = \mathcal{F}_{k} (\bar{\rho}, \bm{\bar{v}},\bm{\bar{B}})$ $(k={\rm I, C, L, V})$ and the contributions of the 
fluctuation components by $\mathcal{F}_{k}^{\rm f} = \mathcal{F}_{k} -\mathcal{F}_{k}^{\rm m}$. Figure 15~(b) illustrates the roles of the 
mean and fluctuation components in the azimuthal force balance. Here we show the radial profiles of 
$\langle\langle \mathcal{F}_{\rm I}^{\rm f}\rangle_s\rangle+\langle\langle \mathcal{F}_{\rm C}^{\rm m}\rangle_s\rangle+\langle\langle\mathcal{F}_{\rm V}^{\rm m}\rangle_s\rangle$ and $\langle\langle \mathcal{F}_{\rm I}^{\rm m} \rangle_s\rangle$ by the red dashed and magenta dashed curves,  
which are all flow origins. The blue and green solid curves denote the radial profiles of 
$\langle\langle\mathcal{F}_{\rm L}^{\rm f} \rangle_s\rangle$ and $\langle\langle\mathcal{F}_{\rm L}^{\rm m} \rangle_s\rangle$, which are 
magnetic field origins. The contributions of fluctuation components to Coriolis and viscous forces are negligibly small. 

In the region where the convective motion is less vigorous (upper convection zone and stable zone), the positive azimuthal force due to the flow field 
is balanced with the negative force sustained by the dynamo-generated magnetic field. In contrast to that, the negative azimuthal force due to the flow field 
is compensated by the positive Lorentz force in the most of the convection zone where the convective motion is vigorous. The Lorentz force by the nonlinear 
coupling of fluctuating magnetic field plays a crucial role in the azimuthal force balance for maintaining the equilibrated profile of the differential rotation. 

As shown in Figure 6, the differential rotation profile established in our model exhibits more cylindrical alignment than the solar rotation profile characterized 
by the conical iso-rotation surface. Our MHD convection system is still dominated by Taylor-Proudmann balance. It is well known that the latitudinal entropy variation at the base of the convection zone induces a baroclinicity, and yields the solar-like conical rotation profile (e.g., \citealt{kitchatinov+95,rempel05, miesch+06,masada11}). It might be important to numerically capture with higher accuracy  the nonlinear MHD processes, such as instabilities and resultant turbulence, in the stable layer to reproduce the large-scale solar convection profile more accurately. 
\subsection{Qualitative Picture of Magnetic Dynamo}
\begin{figure*}[tbp]
\begin{center}
\scalebox{0.8}{{\includegraphics{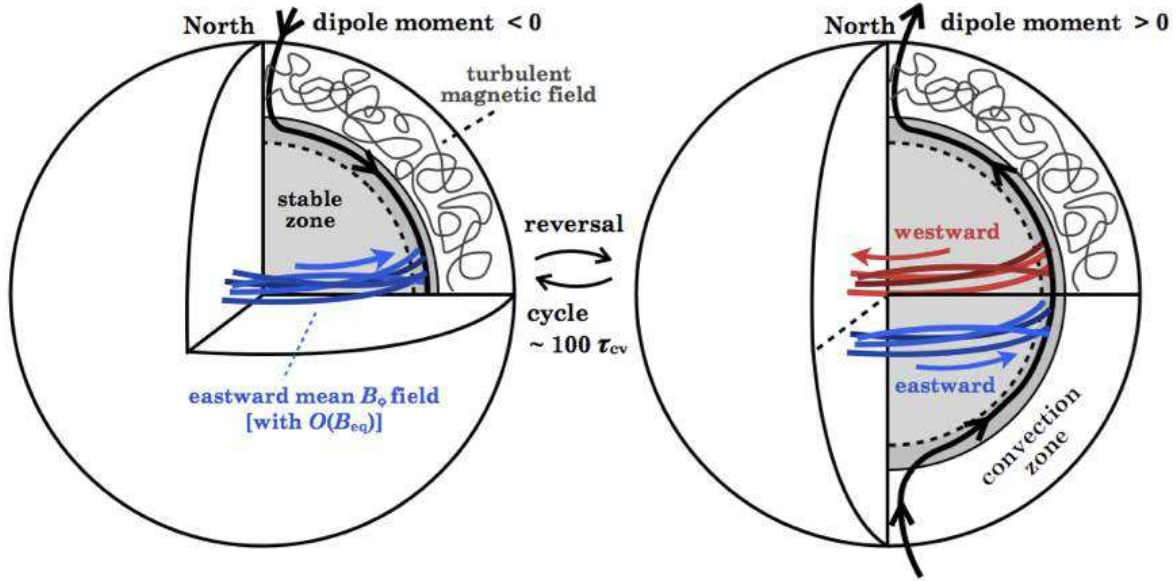}}} 
\caption{Schematic picture of the magnetic structure in the model with the stable layer. The thick black curve demonstrates the poloidal 
magnetic field. The blue and red curves denote the eastward and westward azimuthal magnetic components. The mean large-scale 
magnetic field with $\mathcal{O}(0.1)B_{\rm eq}$ is preferentially organized in the stably stratified layer, whereas the convection zone 
is dominated by the turbulent fluctuating magnetic component. The polarity reversal with the cycle period of $\sim 100 \tau_{c}$ takes 
place globally and synchronously throughout the system.  }
\label{fig16}
\end{center}
\end{figure*}
Since the main purpose of this paper is not to accurately model the solar dynamo, but rather to reveal the effects of the penetrative convection 
on the magnetic dynamo process, the effective luminosity used in the simulation are larger than the solar values. Nevertheless, it would be 
helpful to evaluate the cycle period and the mean-field strength obtained in our simulation in comparing our model with the models of 
other groups. 

Figures~13 gives the cycle period of the polarity reversal $\tau_{\rm cycle} \simeq 100\tau_c$, which is evaluated as
\begin{equation}
\tau_{\rm cycle} \simeq 100\times \frac{d}{v_{\rm rms}} = 6.3\ \ {\rm  [year]} \;,
\end{equation}
when we adopt the solar values $d = 0.3R_\odot$ and $v_{\rm rms} \simeq 100 \ {\rm m\; sec^{-1}}$. This is about a half of the  
cycle of the polarity reversal in the Sun. At the cycle maxima, the strength of the axi-symmetric azimuthal field in the stable zone 
reaches $B_{\rm eq}$, which can be evaluated as
\begin{equation}
B_{\phi,{\rm max}} =  (4\pi \rho_m \; v_{\rm rms}^2)^{1/2} \simeq 8000\ \  {\rm [G]} \;,
\end{equation}
with the density $\rho_{m} \simeq 0.05\ {\rm g\; cm^{-3}}$ at the mid convection zone of the Sun. This is a comparable strength with the 
large-scale magnetic field simulated in \citet{browning+06} and \citet{ghizaru+10}, but would be an order of magnitude smaller than that 
expected in the tachocline layer of the Sun for explaining the sunspot emergence at the surface on 
latitudes of less than $\pm 40^\circ$ (c.g., \citealt{choudhuri+87}) 

When taken all the numerical results together, the structure and evolution of the dynamo-generated magnetic field in our model is represented 
by a schematic picture in Figure~15. The thick black line demonstrates the poloidal magnetic component with dipole dominance. The blue and 
red curves denote eastward and westward azimuthal components. The mean large-scale azimuthal field with $\mathcal{O}(B_{\rm eq})$ 
is preferentially organized around the equatorial region in the stably stratified layer, whereas the convection zone is dominated by the 
turbulent fluctuating component. The polarity reversal with the cycle period of $\sim 100 \tau_{c}$ takes place globally and synchronously 
throughout the system. 

The dipole dominance is one of remarkable features of the large-scale magnetic field rooted in the stable layer. The tendency of the dipole 
dominance that appears when taking account of the stable layer is reported in \citet{browning+06} and \citet{miesch+09}. The differential 
rotation can not only amplify the mean toroidal fields through the so-called $\Omega$-effect, and but also expel the asymmetric field 
components via rotational smoothing process (\citealt{radler80,radler86,spruit99}). The dipole-like magnetic structure with large-scale axi-symmetric 
azimuthal component would be thus a natural outcome of the rotational amplification and smoothing of the magnetic field in the stable layer. 
The more accurate modeling of the stably stratified tachocline would enable us to tackle the generation mechanism of large-scale 
$\mathcal{O}(10^5)\: {\rm G}$ field that can be responsible for the origin of the sunspot on the solar surface. 

We finally remark that both the equatorward migration and buoyant emergence of the large-scale magnetic component, that 
can bridge the gap between the simulation and sunspot observation, could not be simulated in our model. This clearly tells us that we have 
still a lot of missing ingredients to reproduce the solar interior in our dynamo modeling. 

\section{Summary}
We reported, in this paper, our first results of solar dynamo simulation based on the Yin-Yang grid with the fully compressible MHD model.
To investigate influences of the stably stratified layer below the convection zone, two simulation models with and without the stable layer 
(Models A and B) were compared. It is confirmed from our numerical study that the stable layer has substantial influence on the convection 
and the magnetic field. Our main findings are summarized as follows. 

1. The convective motion in the upper convection zone is characterized by upflow dominant cells surrounded by networks of narrow downflow 
lanes for both models. While the radial flow is restrained by the boundary placed on the bottom of the convection zone in Model~B, the downflow 
lanes persist the plume-like coherent structure even just above the bottom of the unstable layer in Model A. The downflow plumes then penetrate 
into the underlying stable layer. 

2. The differential rotation profiles in both models are reasonably solar-like with equatorial acceleration. However, both exhibit more cylindrical alignment 
than the solar rotation profile with the conical iso-rotation surface inferred from the helioseismology. It is remarkable that the radial shear layer, which is 
reminiscent of the solar tachocline, is spontaneously developed without any forcing just beneath the convection zone as a result of the penetrative convection 
in Model A. The Lorentz force by the nonlinear coupling of fluctuating magnetic field plays an important role in the azimuthal force balance for maintaining the  solar-like differential rotation. 

3. While the turbulent magnetic field becomes predominant in the region where the convective motion is vigorous, the mean-field component is preferentially 
built up in the region where the convective motion is less vigorous. Especially in the stably stratified layer, the strong large-scale 
azimuthal component with antisymmetric profile with respect to the equator and the poloidal field with dipole dominance are spontaneously organized. 

4. The mean magnetic component undergoes polarity reversals with the cycle period of $\sim 100\tau_c$ for both models. 
It takes place globally and synchronously throughout the system regardless the presence of the stable layer. However, the amplitude, coherency 
and dipole dominance of the mean magnetic component are much weaker in the model B compared with those in the model A. The stably stratified layer 
is a key component for organizing the large-scale strong magnetic field, but is not essential for the polarity reversal. 

All the dynamo simulations reported here have used a relatively weak stratification with the density contrast of about $3$ (see \S2). The strong stratification 
in the actual Sun may influence on the physical properties of convections, mean flows and magnetic dynamo (\citealt{kapyla+13}). It would be interesting 
that the three key features, solar-like $\bar{v}_\phi$, $\bar{B}_\phi$, and the polarity reversals are self-consistently reproduced, without assuming any 
forcing, even in the modest density stratification. Higher resolution simulations with a more realistic density stratification will facilitate our understanding 
of the physics of the solar convection and the solar dynamo, that is our next step with the Yin-Yang solar dynamo simulation code.

\acknowledgments
We thank the anonymous referee for constructive comments. 
Numerical computations were carried out on $\pi$-Computer at Kobe Univ., K-Computer at RIKEN, and Cray XC30 at National Astronomical 
Observatory of Japan.  This work was supported by JSPS KAKENHI Grant numbers  24740125 and 20260052, and also by the Takahashi 
Industrial and Economic Research Foundation. 
\appendix
\section{Dividing spherical mean of energy into axi-symmetric and asymmetric parts}
The purpose of this section is to split the spherical mean of energy into axi-symmetric part and asymmetric part. 
For a smooth function $h(\theta,\phi)$ on a sphere, we define the following three means. \\
\\
{\it Longitudinal mean}: 
\begin{equation}\label{eq:1334}
\bar{h}(\theta) = \langle h(\theta,\phi)\rangle_\phi :=  \frac{1}{2\pi} \int_{-\pi}^{\pi}\, h(\theta,\phi)\, {\rm d}\phi \;,
\end{equation}
{\it Latitudinal mean}:
\begin{equation}\label{eq:1334b}
\langle h(\theta,\phi)\rangle_\theta :=  \frac{1}{2} \int_{-1}^{1}\, h(\theta,\phi)\, {\rm d}\cos\theta, \;,
\end{equation}
{\it Surface mean}: 
\begin{equation}\label{eq:1334c}
\langle h(\theta,\phi)\rangle_S := \frac{1}{4\pi} \int_{-1}^{1} \int_{-\pi}^{\pi}\, h(\theta,\phi)\, {\rm d}\cos\theta \, {\rm d}\phi = \langle \langle h\rangle_\phi \rangle_\theta
 			     = \langle \langle h\rangle_\theta \rangle_\phi \;.
\end{equation}

\medskip
We can always divide $h(\theta,\phi)$ into axis-symmetric and asymmetric parts:
\begin{equation} \label{eq:1336}
 h(\theta,\phi) = \bar{h}(\theta) + h_a(\theta,\phi).
\end{equation}
Note that
\begin{equation} \label{eq:1337b}
 \langle h_a(\theta,\phi) \rangle_\phi = 0, 
 \quad \langle \bar{h}(\theta) \rangle_S = \langle \bar{h}(\theta)\rangle_\theta.
\end{equation}
The surface mean of $h^2$ is also divided into two parts:
\begin{eqnarray} \label{eq:1338}
 \langle h(\theta,\phi)^2 \rangle_S 
  &=&  \langle (\bar{h}+h_a)^2 \rangle_S \\
  &=&  \langle (\bar{h})^2 \rangle_S 
        + 2  \langle \bar{h} h_a \rangle_S 
         + \langle (h_a)^2 \rangle_S\\
  &=&  \langle (\bar{h})^2 \rangle_\theta + \langle (h_a)^2 \rangle_S \quad\hbox{[ c.f.~eq.~(\ref{eq:1337b}) ]}.
\end{eqnarray}
Expanding $\bar{h}(\theta)$ by the normalized Legendre polynomials
\begin{equation} \label{eq:1348}
  P^\ast_\ell(\cos\theta) = \sqrt{\frac{2\ell+1}{2}} P_\ell(\cos\theta),
\end{equation}
that satisfy
\begin{equation} \label{eq:1348b}
 \int_{-1}^{1} P^\ast_\ell\, P^\ast_{\ell'} \, {\rm d}\cos\theta = \delta_{\ell\ell'},
\end{equation}
as 
\begin{equation} \label{eq:1349}
 \bar{h}(\theta) = \sum_{\ell=0}^\infty H_\ell\, P^\ast_\ell(\cos\theta),
\end{equation}
we get the Perseval's equation,
\begin{equation} \label{eq:1341}
  \langle (\bar{h})^2 \rangle_\theta = \frac{1}{2} \sum_{\ell=0}^\infty H_\ell^2,
\end{equation}
where Legendre coefficients $H_\ell$ are  given by
\begin{equation} \label{eq:1342}
  H_\ell = \int_{-1}^1 \bar{h}(\theta)\, P^\ast_\ell(\cos\theta)\, {\rm d}\cos\theta.
\end{equation}

\medskip
Similarly, for a vector field $\bm{b}(\theta,\phi)=(b_r(\theta,\phi),b_\theta(\theta,\phi),b_\phi(\theta,\phi))$, we get
\begin{eqnarray} \label{eq:1355}
    \langle \bm{b}^2 \rangle_S 
      &=& \sum_{i=\left\{r,\theta,\phi\right\}}  \langle (b_i)^2 \rangle_S\\\label{eq:1423}
      &=& \sum_{i=\left\{r,\theta,\phi\right\}} \left[ \langle (\bar{b}_i)^2 \rangle_\theta
                                                                  + \langle (b_{i,a})^2 \rangle_S
                                                                   \right],\\\label{eq:1431}
   &=&  \hbox{s.p.} + \hbox{a.p.}
\end{eqnarray}
where the symmetric part
\begin{equation} \label{eq:1457}
  \hbox{s.p.} =    \langle (\bar{b}_r)^2\rangle_\theta
         +  \langle( \bar{b}_\theta)^2 \rangle_\theta
         + \langle(\bar{b}_\phi)^2 \rangle_\theta,
\end{equation}
and asymmetric part
\begin{equation} \label{eq:1455}
        \hbox{a.p.} =  \sum_{i=\left\{r,\theta,\phi\right\}}  \langle (b_{i,a})^2 \rangle_S.
\end{equation}
Due to the Perseval's equation~(\ref{eq:1341}), 
each of the three terms in eq.~(\ref{eq:1457}) can be expanded as,
\begin{equation} \label{eq:1415}
  \langle (\bar{b}_i)^2 \rangle_\theta = \frac{1}{2}\sum_{\ell=0}^\infty (\bar{B}_{i,\ell})^2
\end{equation}
where $\bar{B}_{i,\ell}$ are Legendre coefficients
\begin{equation} \label{eq:1417}
  \bar{B}_{i,\ell} = \int_{-1}^1 \bar{b}_i(\theta)\, P^\ast_\ell(\cos\theta)\, d\cos\theta.
\end{equation}
Note that for a magnetic field, the monopole component $\bar{B}_{r,\ell=0}$ is absent.
\\
\\
\\

\clearpage

\end{document}